\documentclass[usenatbib,psfig]{mnras}

\usepackage[T1]{fontenc}

\DeclareRobustCommand{\VAN}[3]{#2}
\let\VANthebibliography\thebibliography
\def\thebibliography{\DeclareRobustCommand{\VAN}[3]{##3}\VANthebibliography}
\usepackage[latin1]{inputenc}
\usepackage{amsmath,amsfonts,amssymb}
\usepackage{color}

\usepackage{times}
\usepackage{url}
\newcommand{\cha}[1]{\textcolor{black}{#1}}

\usepackage{epsfig}
\usepackage{graphicx}
\usepackage{subcaption}
\usepackage[export]{adjustbox}

\usepackage{caption}
\usepackage{changepage}
\usepackage[justification=centering]{caption}
\usepackage{natbib}
\usepackage{wrapfig}

\usepackage{dirtytalk}
\usepackage{amssymb}
\usepackage{placeins}
\usepackage{threeparttable}
\usepackage{color}
\usepackage{float}
\newcommand{\chab}[1]{\textcolor[rgb]{0, 0, 0}{#1}}
\newcommand{\chaS}[1]{\textcolor[rgb]{0, 0, 0}{#1}}
\newcommand{\chabb}[1]{\textcolor[rgb]{0, 0, 0}{#1}}

\usepackage[normalem]{ulem}
\usepackage{dblfloatfix}
\usepackage{multicol}
\usepackage[version=4]{mhchem}
\usepackage{titlesec}

\usepackage[latin1]{inputenc}
\usepackage{amsmath,amsfonts,amssymb}
\usepackage{color}

\usepackage{times}

\newcommand{\Msun}{\ensuremath{\textrm{M}_{\odot}}}

\newcommand{\Lsun}{\ensuremath{\textrm{L}_{\odot}}}

\newcommand{\kms}{km~s$^{-1}$} 

\newcommand{\OI}{\mbox{O\hspace{0.25em}{\sc i}}}

\newcommand{\CII}{\mbox{C\hspace{0.25em}{\sc ii}}}

\newcommand{\MgII}{\mbox{Mg\hspace{0.25em}{\sc ii}}}

\newcommand{\SII}{\mbox{S\hspace{0.25em}{\sc ii}}}

\newcommand{\SiII}{\mbox{Si\hspace{0.25em}{\sc ii}}}
\newcommand{\SiIII}{\mbox{Si\hspace{0.25em}{\sc iii}}}

\newcommand{\CaII}{\mbox{Ca\hspace{0.25em}{\sc ii}}}
\newcommand{\CaIII}{\mbox{Ca\hspace{0.25em}{\sc iii}}}

\newcommand{\FeII}{\mbox{Fe\hspace{0.25em}{\sc ii}}}
\newcommand{\FeIII}{\mbox{Fe\hspace{0.25em}{\sc iii}}}
\newcommand{\CoII}{\mbox{Co\hspace{0.25em}{\sc ii}}}
\newcommand{\CoIII}{\mbox{Co\hspace{0.25em}{\sc iii}}}

\newcommand{\NiIII}{\mbox{Ni\hspace{0.25em}{\sc iii}}}

\newcommand{\Fefs}{$^{56}$Fe}
\newcommand{\Feff}{$^{54}$Fe}
\newcommand{\Cofs}{$^{56}$Co}
\newcommand{\Nifs}{$^{56}$Ni}
\newcommand{\Nife}{$^{58}$Ni}
\newcommand{\Mej}{M$_{\textrm{ej}}$}

\newcommand{\Dm}{\ensuremath{\Delta m_{15}(B)}}

\newcommand{\eg}{e.g.\ }
\newcommand{\ie}{i.e.\ }

\def\gsim{\mathrel{\rlap{\lower 4pt \hbox{\hskip 1pt $\sim$}}\raise 1pt \hbox {$>$}}}
\def\lsim{\mathrel{\rlap{\lower 4pt \hbox{\hskip 1pt $\sim$}}\raise 1pt \hbox {$<$}}}
\def\gtaprx {\lower .1ex\hbox{\rlap{\raise .6ex\hbox{\hskip .3ex
	{\ifmmode{\scriptscriptstyle >}\else
		{$\scriptscriptstyle >$}\fi}}}
	\kern -.4ex{\ifmmode{\scriptscriptstyle \sim}\else
		{$\scriptscriptstyle\sim$}\fi}}}
\def\ltaprx {\lower .1ex\hbox{\rlap{\raise .6ex\hbox{\hskip .3ex
	{\ifmmode{\scriptscriptstyle <}\else
		{$\scriptscriptstyle <$}\fi}}}
	\kern -.4ex{\ifmmode{\scriptscriptstyle \sim}\else
		{$\scriptscriptstyle\sim$}\fi}}}

\usepackage{graphicx}	
\usepackage{amsmath}	
\usepackage{caption}
\title[iPTF16abc]{Abundance stratification in type Ia supernovae -- VII. The peculiar, C-rich iPTF16abc: highlighting diversity among luminous events }

\author[Aouad et al]{Charles J. Aouad$^1$\thanks{E-mail:charlesaouad@aascid.ae}, 
Paolo A. Mazzali$^{1,2}$,
Chris Ashall$^{3}$,
Masaomi Tanaka $^{4,5}$,
Stephan Hachinger$^{6}$
\\
\\
$^1$ Astrophysics Research Institute, Liverpool John Moores University, 146 Brownlow Hill, Liverpool L3 5RF, UK\\
$^2$ Max-Planck Institut f\"{u}r Astrophysik, Karl-Schwarzschild-Str. 1, D-85748 Garching, Germany\\
$^3$ Department of Physics, Virginia Tech, 850 West Campus Drive, Blacksburg, VA 24061, USA\\
$^4$ Astronomical Institute, Tohoku University, Aoba, Sendai 980-8578, Japan\\
$5$ Division for the Establishment of Frontier Sciences, Organization for Advanced Studies, Tohoku University, Sendai 980-8577, Japan\\
$6$ Leibniz Supercomputing Centre (LRZ) of the BAdW, Boltzmannstr. 1, D-85748 Garching, Germany
}

\date{Accepted XXX. Received YYY; in original form ZZZ}

\pubyear{2024}

\begin{document}
\label{firstpage}
\pagerange{\pageref{firstpage}--\pageref{lastpage}}
\maketitle


\begin{abstract}  
Observations of Type Ia supernovae (SNe\,Ia) reveal diversity, even within assumed subcategories. Here, the composition of
the peculiar iPTF16abc \chabb{(SN\,2016bln)} is derived by modeling a time series of optical spectra.
iPTF16abc's early spectra combine traits of SNe 1999aa and 1991T known for weak
\SiII\ $\lambda$ 6355 and prominent \FeIII\ features. However, it differs with  weak early \FeIII\ lines, and persistent \CII\ lines
post-peak. It also exhibits a weak \CaII\ H\&K feature  aligning it with SN\,1991T, an observation supported by their bolometric light curves. The early attenuation of \FeIII\ results from abundance effect.
The weakening of the \SiII\ $\lambda$ 6355 line, stems from silicon depletion in the
outer shells, a characteristic shared by both SNe 1999aa and 1991T, indicating a
common explosion mechanism that terminates nuclear burning at around 12000
\kms\, unseen in normal events. 
\chaS{Beneath a thin layer of intermediate mass elements (IMEs) with a total mass of 0.18 \Msun}, extends a \Nifs\ rich shell totaling 0.76 \Msun\ and generating a bolometric luminosity as high \chaS{as ${L_{\mathrm{peak}}}=1.60 \pm
0.1 \times$ $10^{43}$ ergs s$^{-1}$}.
Inner layers, typical of SNe\,Ia, hold neutron-rich elements, (\Feff\ and \Nife), totaling 0.20 M${\odot}$. Stable iron, exceeding solar abundance, and carbon, coexist in the outermost layers, challenging existing explosion models. The presence of carbon down to $v\approx$ 9000\,\kms, totalling $\sim$ 0.01 \Msun\, unprecedented in this class, links iPTF16abc to SN\,2003fg-like events.
The retention of 91T-like traits in iPTF16abc underscores
its importance in \chaS{understanding} the diversity of SNe\,Ia. 
\end{abstract}



\section{INTRODUCTION}

It is generally accepted that Supernovae type Ia (SNe\,Ia) result from the thermonuclear disruption of a carbon-oxygen (CO) white dwarf (WD) in a 
binary orbit \citep{ Hillebrandt2000,mazzali2007,livio_2018_progenitors_of_Ia}.
The width-luminosity relation (WLR) \citep{Phillips1993}, linking the peak brightness of their light curves to their width, has established them as standardizable candles, unveiling the accelerated expansion of the Universe \citep{Phillips1993,Riess1998,Perlmutter1998}.

The event is thought to initiate when the central temperature reaches $\sim 10^9$~K leading to explosive carbon burning. This ignition can occur through accretion from a non-degenerate companion, [single-degenerate scenario (SD); \citet{whelan_SD_scenario}], or by the violent merger of two WDs [double degenerate scenario (DD) \citet{Iben_1984_DD_scenario}].
 In some cases SNe\,Ia \chaS{may} also originate from a triple system [\citet{kushnir_triple_system_Ia}; for a review see \citet{livio_2018_progenitors_of_Ia}]. 
 \chab{The mass can either attain the Chandrasekhar limit, alternatively, it may exceed it in what is known as a super-Chandrasekhar mass scenario, or it may eventually fall below it in a sub-Chandrasekhar mass scenario.} In all cases, a burning front propagates, unbinds the WD and causes the production of heavy elements \chaS{including} radioactive \Nifs\,. The decay of \Nifs\ through the \Nifs~$\rightarrow$~\Cofs~$\rightarrow$~\Fefs\  chain \citep{pankey1962PhDT, Colgate_Mckee_1969_nifsdecay} releases $\gamma$-rays and positrons which deposit energy in the expanding ejecta,  giving rise to optical photons. These photons remain trapped until they diffuse as the ejecta expand and the opacity decreases. \Nifs\ and other iron group elements (IGEs) have a rich array of line transitions. Their presence in the H poor ejecta where line opacity dominates \citep{PauldrachNLTElineblocking}, increases the diffusion time, so events with more \Nifs\ are not only hotter and brighter, but decline slower leading to the observed \chaS{Width-Luminosity relation in their light curves} \citep{Arnett1982, mazzali2001, LC_SNIa_Bersten_Mazzali_2017}.

 It has been suggested that the observed scatter in their peak luminosity corresponds to different amounts of \Nifs\ synthesised 
(a range of 0.1 -- 1 \Msun\,, \citealt{mazzali2001,Mazzali2008}). What causes this spread in \Nifs\ production is still not clear, but it \chaS{is} a direct consequence of the burning regime. 
Currently, \chaS{one of} the favored burning \chaS{scenarios} is one in which a deflagration wave (subsonic burning) transits to a detonation wave (supersonic shock wave) in the so-called deflagration to detonation transition model (DDT; \citealt{khokhlov91DD, mazzali2007}). This transition enables \chab{efficient burning in the} outer layers of the ejecta, resulting in the synthesis of sufficient amounts of intermediate-mass elements (IMEs; \eg\ Si, S, Ca) as supported by observation\chaS{s}. 

A significant proportion of SNe Ia are typically categorized as "normal" owing to the uniformity of their spectroscopic characteristics. Nevertheless, an increasing number of events with a wide array of spectral features have been observed and classified as "peculiar". These peculiar SNe Ia encompass a broad range, from cooler, fainter objects like SN\,1991bg \citep{mazzali91bg}, SN\,1986G \citep{Ashall2016}, SN\,2002cx \citep{Li_2003_2002cx}, SN\,2002es \citep{Ganeshaligam_2002es} to hotter, more luminous objects such as SN\,1991T \citep{mazzali1995} \chaS{or} SN\,1999aa \citep{garavini2004}.
Furthermore, within the realm of luminous SNe\,Ia, there exists a distinct subgroup known as SN 2003fg-like events \citep{ashall_2021_2003fg_superchandra}, which are also referred to as super-Chandrasekhar mass events (SC). (\eg SN\,2003fg -- \citealt{SN2003fg}, SN\,2006gz -- \citealt{hicken_2006gz}, SN\,2012dn -- \citealt{chakdrahari_2012dn}, SN\,2009dc -- \citealt{taubenberger_2011_2009dc}). Yet, despite several attempts to classify their spectra \citep[\eg][]{nugent1995, benetti2005, Branch_classification_2006, wang_classifc_2009}, 
a consensus regarding this observed diversity remains elusive.

During the initial stages of a SNe Ia event \chaS{observed in the optical,} known as the photospheric phase, the spectra probe the outer layers. They are characterized by broad P-Cygni profiles indicating expansion velocities of the order of $ 10,000 $ \kms.  
At this early stage, the main spectral features of the \chaS{SNe} classified as "\textit{normal}" are lines of singly ionized IMEs such as \SiII, \SII, \CaII,   \MgII\ and neutral elements such as \OI\,. As time progresses, deeper layers of the ejecta are revealed and blends of \FeIII\ and \FeII\ start to influence the 
appearance of the spectra. \citep{filippenko_1997_optical_spectra_of_Sn,parrent2014_review_of_Ia_SN_spectra}.

A few months after reaching maximum light, the ejecta \chaS{of SNe\,Ia generally} become gradually transparent to gamma-rays, allowing only a fraction of the radioactive energy decay to be deposited \citep{ashall_2021_2003fg_superchandra}. Around one year after the explosion, SNe\,Ia spectra probe the innermost shells of the ejecta and the supernova reaches the so-called nebular phase, whose spectra are characterized by emission, mostly in forbidden [\FeIII] and [\FeII] lines \citep{mazzali20152011fe}. 

In contrast to normal Type Ia SNe, the early spectra of the events classified as 1991T-like display distinct behavior. Singly ionized IMEs like \SiII\, $\lambda$ 6355, \SII, \CaII\ H\&K, and \CaII\ NIR are either absent or notably weak, while prominent \FeIII\ lines dominate.  As the events progress, their spectra gradually transition to resemble those of normal SNe\,Ia around peak brightness. 
While most 1991T-like events belong to the shallow silicone (SS) sub-class  according to the classification scheme of \citet{Branch_classification_2006} or to the low velocity gradient (LVG) sub-class according to \citet{benetti2005}, substantial diversity exists within this group. 
For example, the initial spectra of SN 1999aa exhibit \CaII\ HK and \SiII\ $\lambda$ 6355 lines with strengths intermediate between those seen in "normal" events and those classified as 91T-like, morphing to resemble the normal events earlier than SN\,1991T \citep{garavini2004, Aouad_99aa}.
Conversely, the early spectra of the events classified as 2003fg-like showcase numerous deep singly ionized IME lines, despite many exhibiting high luminosities. They are also characterized by slow photospheric velocities, and relatively weak \FeIII\, lines. Additionally, they all exhibit strong and persistent carbon features, the signature of unburnt material in the ejecta, observable until post-maximum epochs. 

\begin{figure}
\includegraphics[trim={0 2 0 15},clip,width=0.5\textwidth]{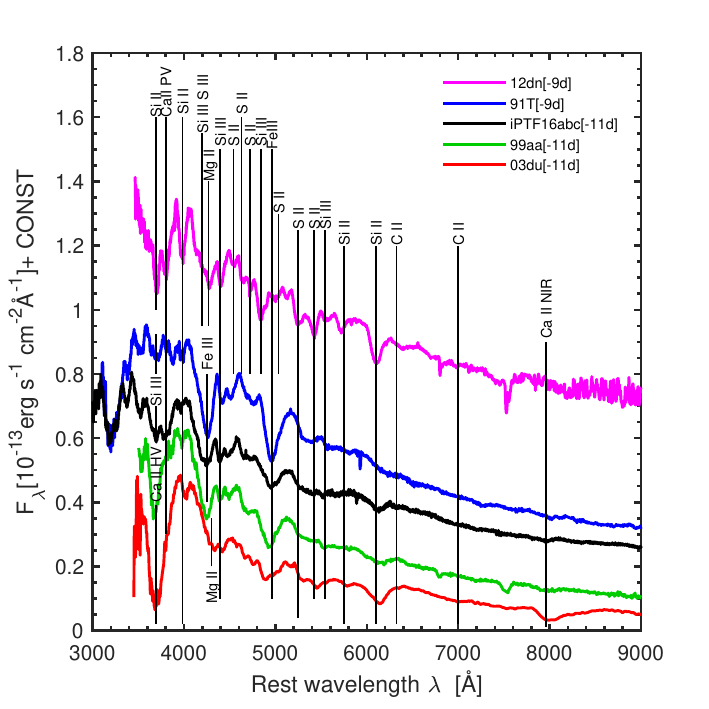}
 \caption{Early time spectra of iPTF16abc in comparison with SN\,1991T, SN\,1999aa, the normal SN\,2003du and the 03-fg like SN\,2012dn. Epochs are shown with reference to $B$ maximum.}  
\label{fig1}
\end{figure}

\begin{figure}
\includegraphics[trim={0 2 0 15},clip,width=0.5\textwidth]{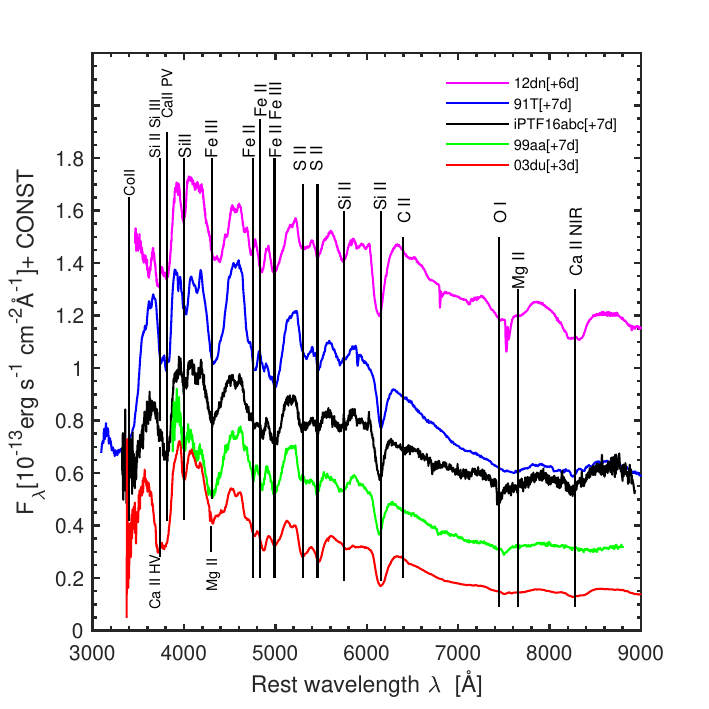}
 \caption{One week post maximum spectrum of iPTF16abc in comparison with SN\,1991T, SN\,1999aa, the normal SN\,2003du and the 03-fg-like SN\,2012dn. Epochs are shown with reference to $B$ maximum.}  
\label{fig2}
\end{figure}

\begin{figure}
\includegraphics[trim={0 2 0 15},clip,width=0.5\textwidth]{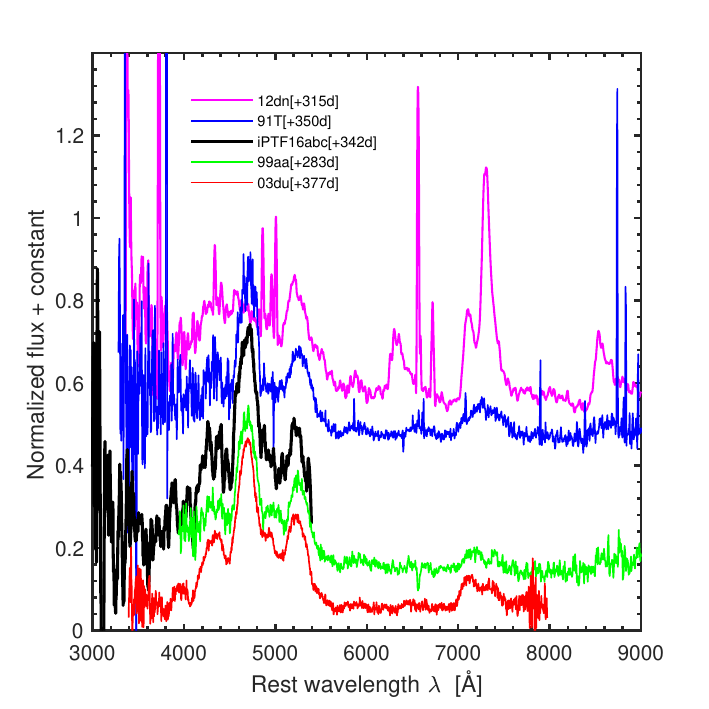}
 \caption{Nebular spectrum of iPTF16abc in comparison with SN\,1991T, SN\,1999aa, the normal SN\,2003du and the 03-fg-like SN\,2012dn. Epochs are shown with reference to $B$ maximum.}  
\label{fig3}
\end{figure}

This spectral diversity challenges our understanding of these objects and their reliability as precise distance indicators. This is especially important for the bright events, which are of particular significance in cosmology. Therefore, it is important to identify objects whose spectroscopic features \cha{link} between two or several different spectroscopic classes. iPTF16abc is an example of such an object. It is characterized by early spectra resembling SN\,1999aa, however, it shows differences when compared to both SN\,1991T and SN\,1999aa (cf. Fig. \ref{fig1}). In particular, the \CaII\ H\&K high velocity (HV) feature is of intermediate strength between SN\,1999aa and SN\,1991T, Fe lines are weaker, and it exhibits a \CII\, $\lambda$$\lambda$ 6578, 6583 line that slowly fades one week before $B$ maximum, \chaS{but} then re-appears one week post maximum. As we progress to one week past peak brightness and extend into the nebular phase, the spectra of these three events become nearly indistinguishable (cf. Figs. \ref{fig2} and \ref{fig3}). 
Remarkably, \chaS{such properties are not completely different from those observed} in 2003fg-like events. \chaS{All} this establishes iPTF16abc as a captivating event, highlighting the subtle spectroscopic diversity that exists within the realm of luminous peculiar events.

In this paper, we employ abundance tomography as a method to investigate SNe\,Ia, allowing us to map the internal composition of the expanding ejecta and gain insights into the explosion process \citep{Stehle2005,Mazzali2008,Tanaka2011,Sasdelli2014,Ashall2016, Aouad_99aa}. We apply this technique to analyze iPTF16abc \chaS{and compare it to} SN\,1999aa, SN\,1991T, and the normal SN\,2003du. While we draw some parallels with 2003fg-like events, a comprehensive exploration of this aspect is reserved for future investigations.
One of the main strengths of our work is that we \chaS{have been applying} a similar modeling technique to analyze and compare different events. This increases confidence that the similarities and differences identified between different SNe\,Ia are real and  not limited by the modeling technique. The paper's structure is as follows: Sections \ref{sec2} and \ref{sec3} discuss data and methods, Sections \ref{sec4} and \ref{sec5} present models for photospheric phase spectra, Section \ref{sec6} focuses on the nebular spectrum model, Section \ref{sec7} discusses the abundance stratification while Section  \ref{sec8} outlines a light curve model based on our abundance stratification results. Section \ref{sec9} discusses our findings, and Section \ref{sec10} concludes the paper.

\vspace{-20pt}

\setlength{\belowcaptionskip}{-1pt}

	\begin{figure}
\includegraphics[trim={15 0 15 0},clip,width=0.5\textwidth]{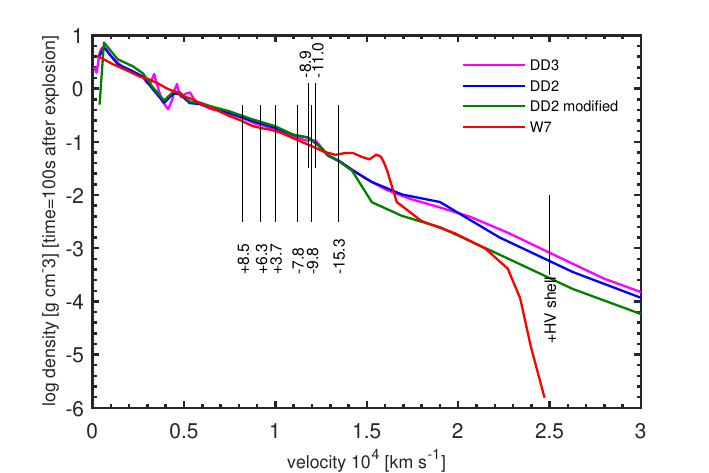}
 \caption{Density profiles used in the modeling process: W7 \citep{nomoto1984}, DD2, DD3 \citep{iwamoto1999}, and the modified DD2. In the modified DD2, a mass of 0.05 M$_\odot$ has been removed at $v$ > 14,000 \kms and redistributed in the inner layers to conserve the total mass. The total Kinetic energy is not significantly changed due to the small mass shift.  Vertical dashed lines mark the photospheres of the synthetic spectra.}  
\label{fig4}
\end{figure}

\section{DATA}
\label{sec2}

iPTF16abc (IAU name: SN 2016bln), was discovered by the intermediate Palomar Transient Factory on April 4, 2016, at 09:21:36.000, JD=2457482.89 \citep{iptfdiscovery}, in the tidal \chab{tail} of NGC 5221, a barred galaxy classified as SABb at a redshift \chaS{of} 0.023279 \citep{redshift_iptf16abc}. Reported distance moduli, $\mu$, to the host galaxy, using the Tully-Fisher relation,  vary from 34.69 to 35.09\textsuperscript{1}\citep{willick_distance_to_NGC5221, Karachentsev_distance_to_NGC5221, distance_to_ngc_5221, hyperleda}. We adopt an extinction value $E(B-V)$ of 0.05
\footnotetext[1]{\url{http://leda.univ-lyon1.fr/}}  for the Milky Way \citep{ferretti2017_probingdustinNGC5221_using_iPTF16abc} and 0.03 for the host galaxy \citep{schlafly2011_reddening, Miller_2018_photometry_of_iptf16abc}. 
Early time photometric data in the $U$, $B$, $V$, $r$ and $i$ bands were taken from \citet{Miller_2018_photometry_of_iptf16abc} and late time $g$ band data were taken from \citet{Dhawan_neb_spect_iptf16abc} who reports a \Dm\, of 0.91. Spectroscopic observations of iPTF16abc in the photospheric phase were obtained with a variety of telescopes and instruments. A nebular spectrum was obtained 342 days after $B$ \chabb{maximum} on 2017-03-29 by \citet{Dhawan_neb_spect_iptf16abc}  .\chabb{The time of the $B$-band maximum light was considered to be on Modified Julian Date (MJD) 57499.54, as adopted in \citet{Miller_2018_photometry_of_iptf16abc}}. Photo-calibration was performed in the $U$, $B$, $V$, $r$, bands by multiplying the spectra with a low-order smoothed spline. The calibrated flux does not \chaS{deviate} by more than 10 percent \chaS{from} the interpolated photometry data in any pass band. The spectra used in the current study are listed in Table \ref{tab1}.

\vspace{-10pt}

\section{MODELLING TECHNIQUES OF THE PHOTOSPHERIC PHASE}
\label{sec3}

We use a 1-D Monte Carlo radiative transfer code \chaS{as laid out in} \citet{Mazzali1993,Lucy1999a,Lucy1999, Mazzali2000, Stehle2005}. The code computes the radiative transfer through the expanding ejecta at a given epoch. It assumes a sharply defined spherical photosphere from which a continuous black body radiation with temperature $T_{\mathrm{ph}}$ is emitted. Seconds after the initial explosion, the ejecta is assumed to \chabb{coast} on a homologous expansion in which the radius is proportional to the velocity \ie $r=v \times t$, therefore the velocity of the photosphere $v_{ph}$ can be used as a spatial coordinate assuming a time $t$ from the explosion. As the ejecta expand, the photosphere recedes in mass coordinates. Photons emitted at this photosphere propagate through the ejecta and undergo Thomson scattering or intersect with the lines assuming the Sobolev approximation \citep{sobolev1960, castor1970}. Bound-bound emissivity is treated through a branching scheme \citep{Mazzali2000}. Ionization and excitation levels are calculated assuming a nebular approximation \citep[cf.][]{Mazzali2000}. In this approximation the gas state \chaS{depends strongly on a} radiation temperature $T_{\mathrm{R}}$, computed from the  mean frequency of the radiation field at every radial mesh, and a dilution factor which parametrizes the radiation field energy density as a function of radius. The radiation field and the gas state are \chaS{iterated} until $T_{\mathrm{R}}$ is converged to the percent level. $T_{\mathrm{ph}}$ is automatically adjusted  to match the given $L_\mathrm{{bol}}$ taking in consideration the back-scattering rate.
The code requires an initial fixed density profile of the SN ejecta, in addition to the input parameters $v_{\mathrm{ph}}$, $L_{\mathrm{bol}}$, $t$ and the abundances in every shell which are \chaS{manually} optimized to match the observed data.  For a detailed description of the \chaS{modelling procedure} see \citet{Stehle2005} and \citet{mazzali_2014_2011fe}, \chab{and for a discussion on errors see} \citet{ashall2020error}.


		
	

\begin{table}
\setlength{\tabcolsep}{2pt}
\caption{Spectra of SN\,iPTF16abc and modelling parameters.}
\label{tab1} 


\scalebox{0.95}{
\hskip-0.5cm\begin{tabular}{llcccccc}

\hline
   UT Date   &  JD$^a$ & Epoch$^b$ & Telescope/Instr. & log\,$L$ & $v$ & $T_\mathrm{ph}$ \\
    2016      &          &   (days)         & &[\Lsun] & (\kms) & (K) \\
  \hline
 05/04$^c$ & 483.3 & $-$15.3  & Gemini-N / GMOS & 8.860 & 13,450 & 15900 \\
 10/04$^c$ & 488.4 & $-$11.0  & \chabb{K}eck-I / LRIS & 9.370 & 12,200 & 14670 \\
 11/04$^c$ & 489.5 & $-$9.8  & LCO-2m / FLOYDS  & 9.470  & 12,000 & 14935\\
 12/04$^c$ & 490.4 & $-$8.9 &  LCO-2m / FLOYDS  & 9.550 & 11,800 & 14832 \\
 13/04$^c$ & 491.5 & $-$7.8 & LCO-2m / FLOYDS  & 9.580 & 11,200 & 15330 \\
 25/04$^c$ & 503.3  & $$+$$3.7  & LCO-2m / FLOYDS   & 9.620 & 10,000 & 10187 \\
 28/04$^c$ & 506.0 & $$+$$6.3  & NOT / ALFOSC  & 9.610 & 9,200 & 10040 \\
 30/04$^c$ & 508.3 & $$+$$8.5 &  LCO-2m / FLOYDS  & 9.520 & 7,500 & 9682 \\
 29/03/2017$^c$ & 841.7 & $$+$$342.4  & \chabb{K}eck-I / LRIS  &   &   &  \\

\hline
\end{tabular}}

\small
$^a$ JD$-$2,457,000 \\
$^b$ Rest-frame time since $B$ maximum \\
$^c$ see \cite{Dhawan_neb_spect_iptf16abc}\\
\end{table}

\section{THE PHOTOSPHERIC PHASE}
\label{sec4}
\begin{figure*}
 
\captionsetup[subfigure]{labelformat=empty}
	\centering
	\begin{subfigure}{1.02\textwidth}
		\includegraphics[trim={10 0 20 20},clip,width=1.02\textwidth]{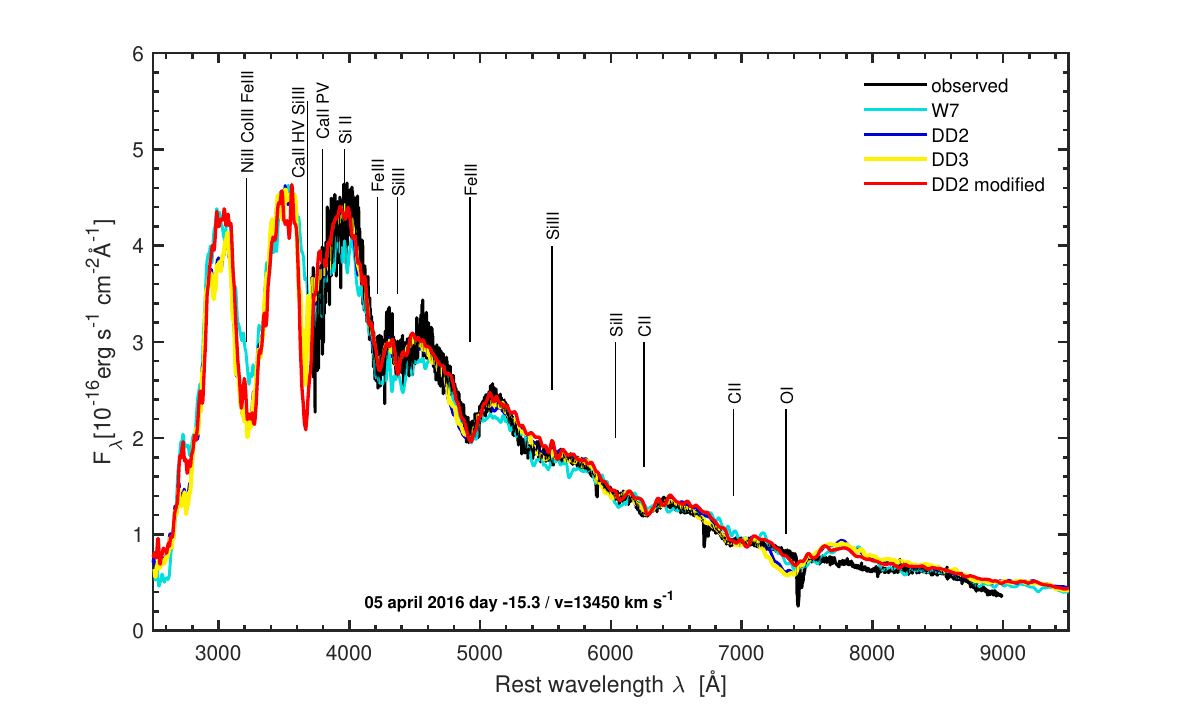}
		\caption{ }
	\end{subfigure}
	\begin{subfigure}{1.02\textwidth}
		\includegraphics[trim={10 0 20 20},clip,width=1.02\textwidth]{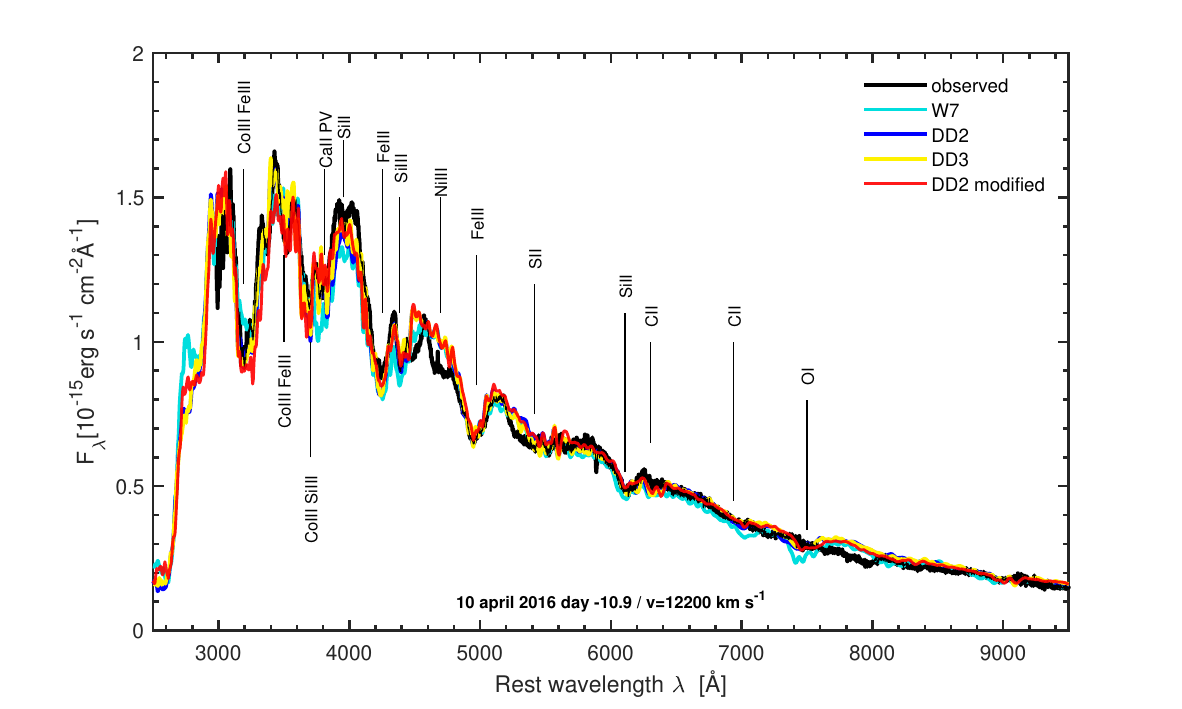}
		\caption{ }
	\end{subfigure}
 	\caption{ Observed early spectra  of iPTF16abc (black), compared to the synthetic spectra in colors, corresponding to different density profiles as indicated in the legend. Epochs are shown with reference to $B$ maximum.
	}
 \label{fig5}
\end{figure*}

\begin{figure*}
 
\captionsetup[subfigure]{labelformat=empty}
	\centering
	\begin{subfigure}{1.02\textwidth}
		\includegraphics[trim={10 0 20 20},clip,width=1.02\textwidth]{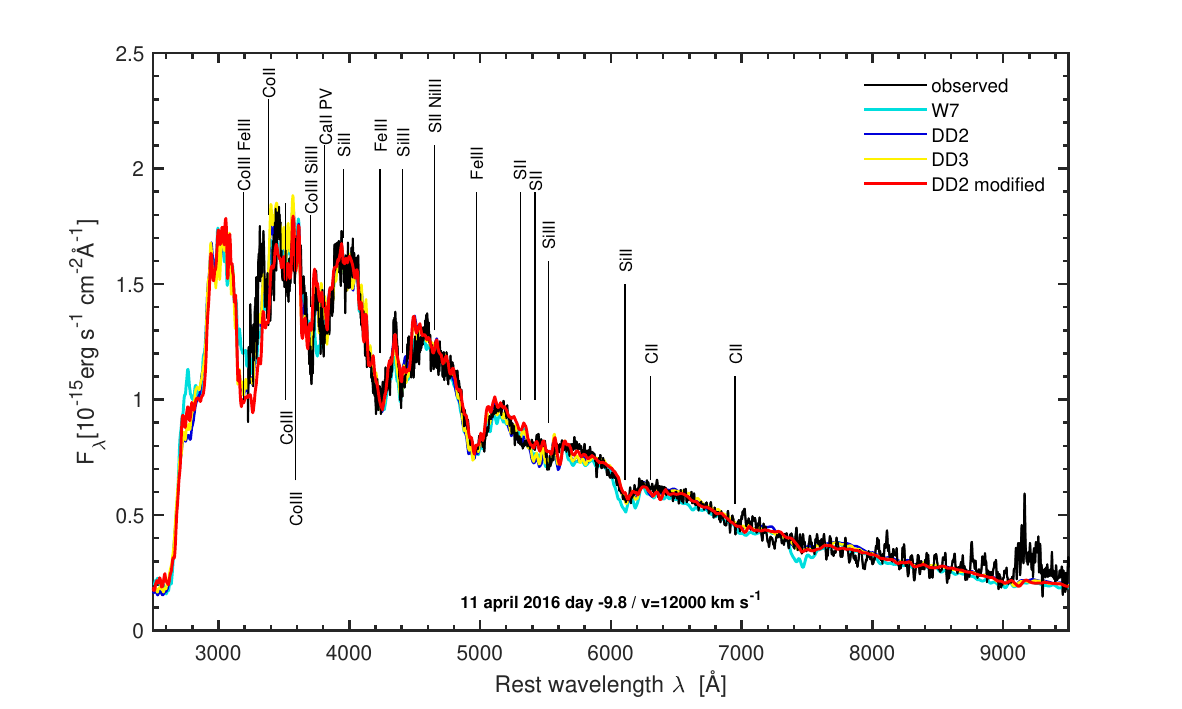}
		\caption{ }
	\end{subfigure}
	\begin{subfigure}{1.02\textwidth}
		\includegraphics[trim={10 0 20 20},clip,width=1.02\textwidth]{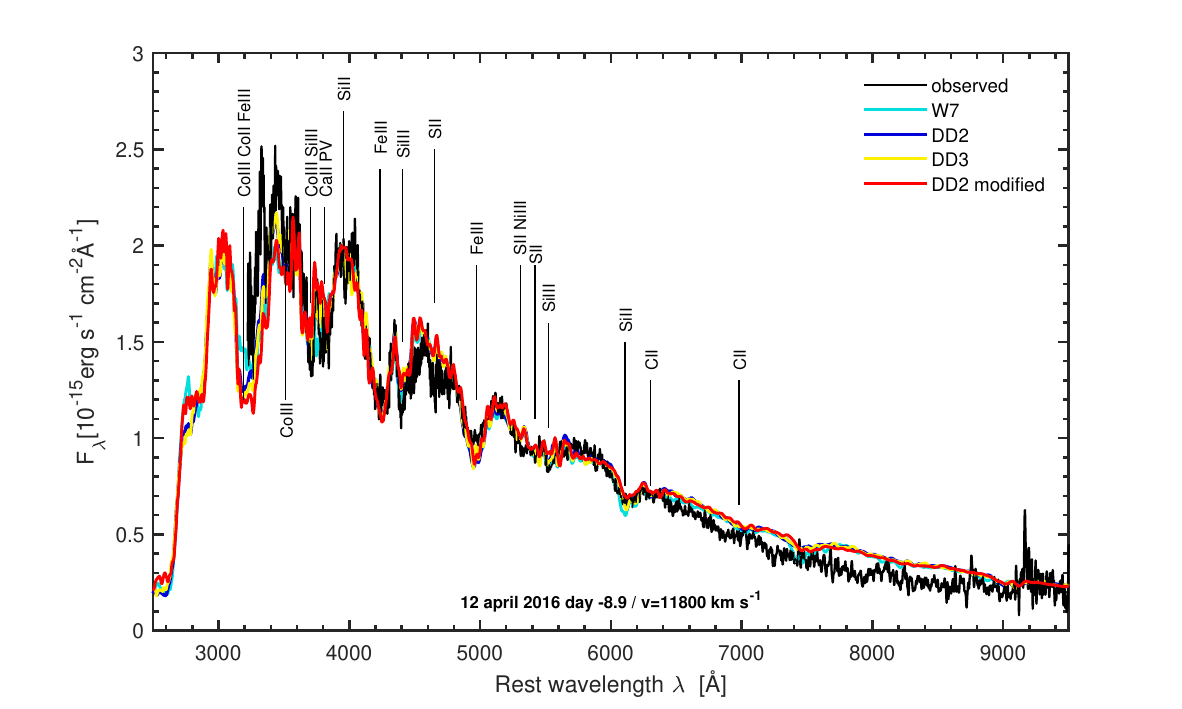}
		\caption{ }
	\end{subfigure}
 	\caption{ Observed early spectra  of iPTF16abc (black), compared to the synthetic spectra in colors, corresponding to different density profiles as indicated in the legend. Epochs are shown with reference to $B$ maximum.
	}
 \label{fig6}
\end{figure*}

\begin{figure*} 
\includegraphics[trim={10 0 20 10},clip,width=1.02\textwidth]{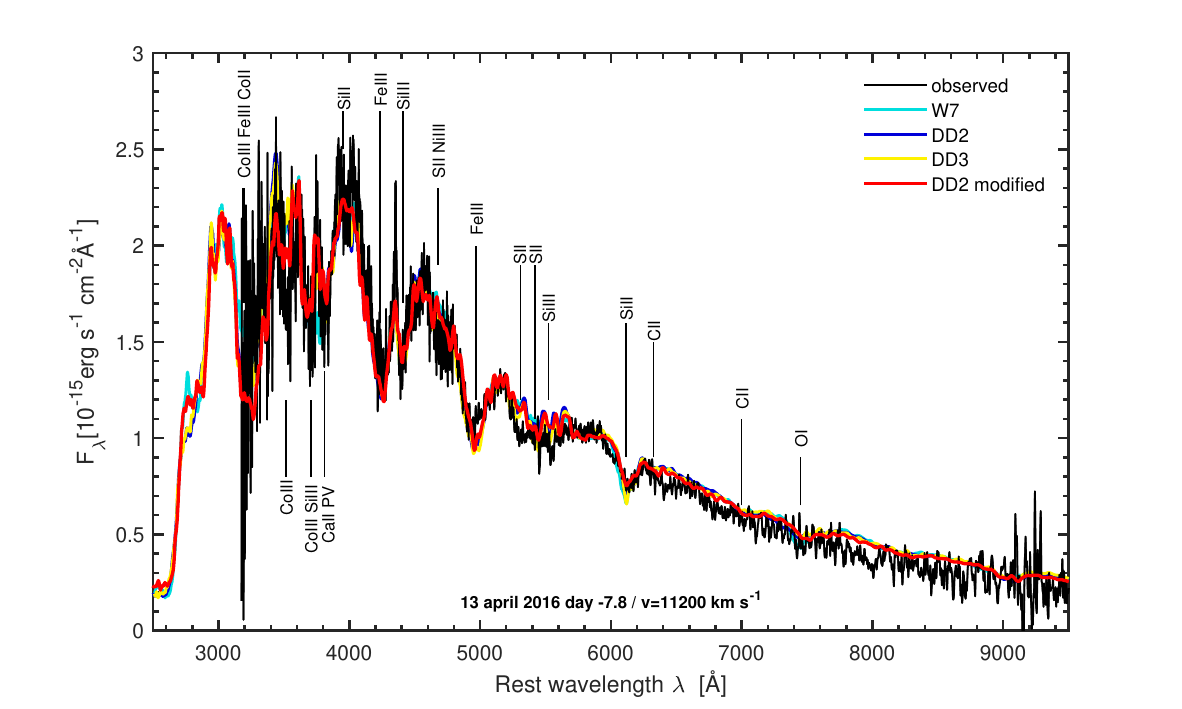}
 \caption{Observed day $-$8 spectrum  of iPTF16abc (black), compared to the synthetic spectra in colors, corresponding to different density profiles as indicated in the legend. Epoch is shown with reference to $B$ maximum.}  
\label{fig7}
\end{figure*}

We analyzed 8 spectra obtained from 15 days before \chaS{to} 8 days after $B$ maximum. To model these spectra, we employed \chabb{four} distinct density profiles, namely one \chab{fast-deflagration} (W7), two delayed detonation models (DD2 and DD3, \citealt{nomoto1984, iwamoto1999}) \chabb{and one modified DD2 density profile (see section \ref{sec5})}; \chabb{The density profiles are shown in} Fig. \ref{fig4}. \chabb{W7, DD2 and DD3}, correspond to well-known hydrodynamic models that have been successfully used to model a number of SNe\,Ia and reproduce the majority of their observed spectroscopic features, both during the photospheric and the nebular phases as well as their light curves \citep{Stehle2005, Mazzali2008,  Tanaka2011, Ashall2016,ashall2007on2011iv, Aouad_99aa}. Using the same density profiles when modelling different supernovae enables comparison between different events, as is the focus of this series of papers. The input parameters used in the models are shown in Table \ref{tab1}, and the resulting spectra are presented in Figures \ref{fig5}, \ref{fig6}, \ref{fig7}, \ref{fig14} and \ref{fig15}.


\subsection{The pre-maximum spectra}
\begin{figure*}
\includegraphics[trim={25 0 25 0},clip,width=0.99\textwidth]{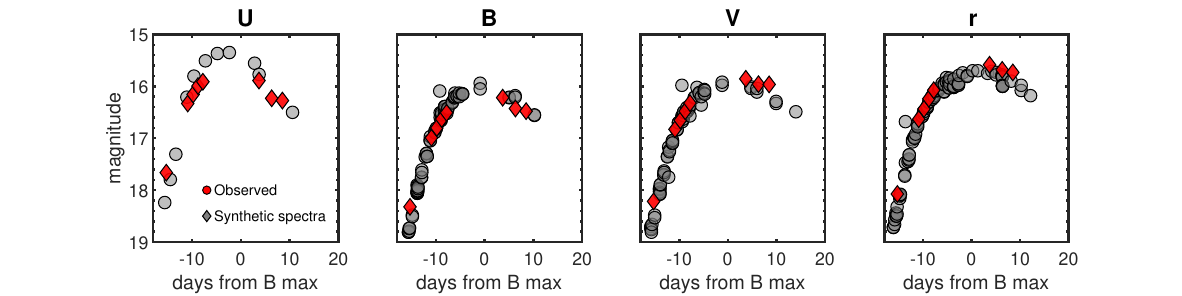}
 \caption{Light curves from observation compared to the ones from synthetic spectra. Photometry data from \citet{Miller_2018_photometry_of_iptf16abc}.}  
\label{fig8}
\end{figure*}

In Figures \ref{fig5}, \ref{fig6}, and \ref{fig7}, we display synthetic spectra from day $-$15 to day $-$8, overlaid on the observed spectra.
Our synthetic spectra faithfully replicate the observed features, with a good match in flux across the entire wavelength range. This alignment is evident in Fig. \ref{fig8}, where we compare photometric light curves derived from synthetic spectra with observational measurements.   
The luminosity exhibits a steep rise from day $-$15 to day $-$11, while the velocities decline slowly to preserve the optimal temperatures required for the synthetic spectra's ionization equilibrium. Notably, the velocities closely resemble those calculated for SN\,1999aa, but they significantly \chab{exceed} those of SN\,2003du, signaling a substantially higher level of opacity in \chaS{SN\,1999aa and iPTF16abc}.

\begin{figure}
\includegraphics[trim={0 0 0 15},clip,width=0.5\textwidth]{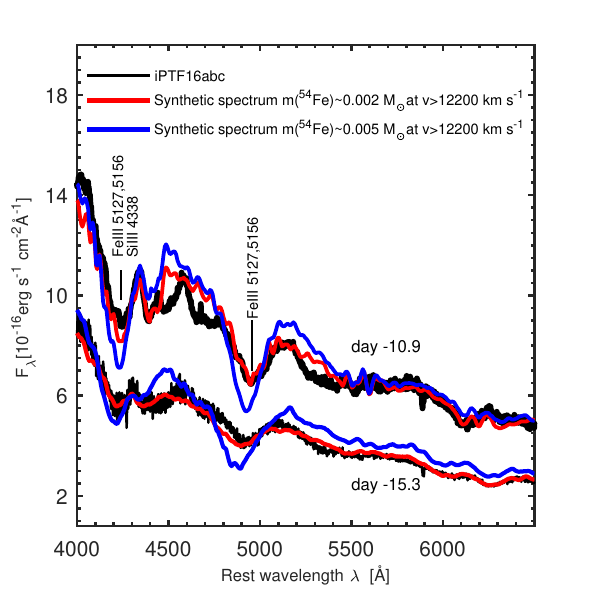}
 \caption{Probing iron at the outermost layers. The blue spectra are models generated replicating an iron abundance similar to SN\,1999aa [X(Fe)=0.018 at $v$ > 12,200 \kms, with a total mass of 0.005 \Msun\ \citep{Aouad_99aa}]. The red spectra have been produced using X(Fe)= 0.00035 at v> 13,450\kms, and X(Fe)=0.025 in the shell above 12,200 \kms, with a total mass of 0.002 \Msun. Notably, despite the outermost shells of iPTF16abc exhibiting super-solar abundances, its iron content in these layers remains lower than that of both SNe 1999aa and 1991T. \chabb{Epochs are shown with reference to $B$ maximum.}}  
\label{fig9}
\end{figure}

\textit{Fe-group elements:} The earliest spectrum of iPTF16abc, taken at day $-$15, shows \FeIII\, features around 4250 and 5000 \AA\, that we could reproduce with X(Fe)=0.0035 at $v>13,450$\,\kms. This is above solar \chab{abundance}
\citep[$X$(Fe$_\odot$) = 0.001,][]{asplund-2009-solar-abundance}. As we advance to day $-$11, these features grow in intensity, \chaS{corresponding to} an increase in the iron mass fraction to 0.025 at $v>12,200$\kms\,.  However, when compared to SNe 1999aa and 1991T at the same epoch (Fig. \ref{fig1}), these features appear comparatively weaker. Consequently, the iron abundance in iPTF16abc remains noticeably lower (Fig. \ref{fig9}). 
The iron present must be stable (\ie \Feff), as it could not have resulted from the decay of \Nifs, at this early stage. Remarkably, the presence of iron at the outermost shell \chab{up to} $v$= 25,000 \kms\, is not anticipated by any of the explosion models.

These features keep increasing in strength at days $-$10 and in the following days. Subsequently, the \chab{Fe} mass fraction gradually increases, reaching a value of 0.038 at day $-$3.8 at $v > 11,200$\,\kms\,. At this epoch, this fraction is only $\sim$ 50 percent of all iron needed, the remaining  being the product of \Nifs\ decay. These stable iron mass fractions values significantly surpass solar abundances, strongly indicating that at these shells ($v$ < 13,450 \kms\,), iron likely results from explosive nucleosynthesis. \chaS{These stable iron mass fractions are similar to the ones computed in both the W7 and the DD models \citep[cf.][]{nomoto1984, iwamoto1999}}. 

\begin{figure} 
\includegraphics[trim={0 0 0 0},clip,width=0.5\textwidth]{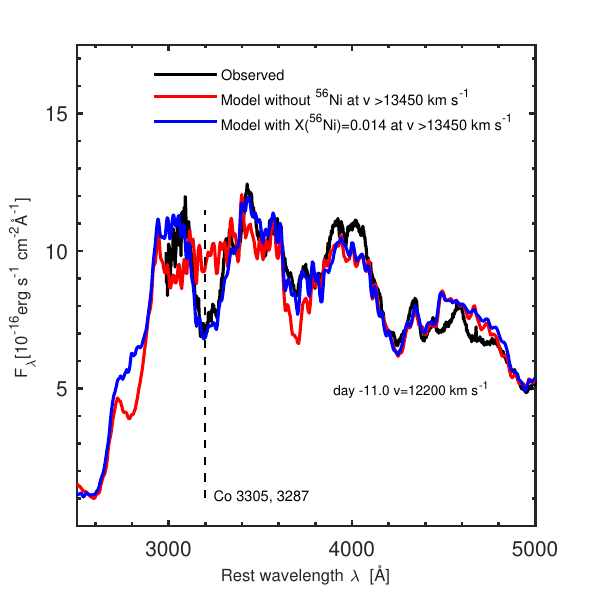}
 \caption{Probing nickel at the outermost layers. A \Nifs\ mass fraction of 0.014 is required at the outermost shell at v > 13,450 \kms to replicate the \CoIII\ $\lambda$$\lambda$ 3287, 3305 line observed near 3200 \AA\,. \chabb{Epoch is shown with reference to $B$ maximum}. }  
\label{fig10}
\end{figure}

A \Nifs\, mass fraction of 0.013 is required at $v>13,450$\kms, to accurately replicate the \CoIII\, $\lambda$$\lambda$ 3287, 3305 feature, observed  near 3200 \AA, in spectra taken on day $-11$ and onwards (cf. Fig. \ref{fig10}). Additionally, this \Nifs\, mass fraction \chaS{suppresses} the flux in the ultraviolet (UV) range and \chaS{contributes to redirecting it} towards longer wavelengths in the earliest spectrum obtained on day $-15$.

Going inwards, the \Nifs\, mass fraction increases gradually, reaching a peak of 0.69 at $v > 11,200$\,\kms at day $-$8. These values are slightly larger than the ones computed for SNe 1999aa and 1991T at the same epochs. 
A feature centered around 4700 \AA, primarily influenced by multiple \NiIII, and \SII, lines is also replicated in the synthetic spectra.  
Unfortunately, a comparison  with SNe 1999aa and 1991T is not possible at day $-$15 as data for these events are not available at this early stage. 

\begin{figure}
\includegraphics[trim={0 0 0 0},clip,width=0.4\textwidth]{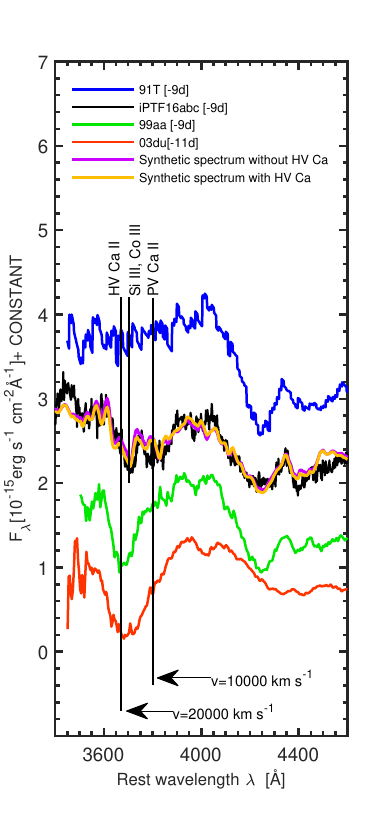}
 \caption{Detail of the region around $\sim 3700$ \AA.  the HV \CaII\ feature reveals a split into two components. The bluer portion is characterized by the influence of \SiIII\ and \CoIII\ rather than \CaII\ in iPTF16abc. Interestingly, a synthetic spectrum at day -9, lacking calcium at high velocity (X(Ca)=0> 13,450 km/s), shows no discernible difference in its bluer segment compared to the one featuring Ca.
In contrast, in SN\,1999aa, \CaII\ dominates the same feature at the same time, and this dominance is even more pronounced in SN\,2003du. \chabb{Epochs are shown in square brackets with reference to $B$ maximum.}}  
\label{fig11}
\end{figure}

\textit{Calcium:} 
In the majority of SNe\,Ia, the \CaII\ H\&K feature, seen around 3800 \AA\, is often composed of two separate components: one at photospheric velocity (PV) and one at high velocity (HV) \citep{dutta_2017hpa_decomposition_of_ca_lines, childress2014_HVfeatures, mazzali2005HVfeatures}. In the SNe Ia classified as spectroscopically normal, \eg SN\,2003du  \citep{Tanaka2011}, the HV component dominates, causing the two features to blend together and form a single deep absorption line. However,  the high-velocity component in iPTF16abc  at day $-$11 and after, 
is relatively weak, causing the feature to split into two separate components with the blue part being of intermediate strength between the one seen in SN\,1999aa and the one seen in SN\,1991T (cf. Fig. \ref{fig1}).
 Our synthetic spectra accurately reproduce this behavior with X(Ca)=0.00055 at $v > 25,000$\,\kms, which is above Solar abundance \citep[$X$(Ca$_\odot$) = 0.00006 as reported in][]{asplund-2009-solar-abundance}. 
 This fraction decreases to 0.0002 at $v > 13,450$ \kms than increases to 0.01 at $v > 12,200$ \kms and remains constant until day $-$9 at $v > 11,800$ \kms. 
 This calcium mass fraction at lower velocities is necessary to be able to replicate the (PV) component of the feature. Even with this increase in the calcium mass fraction, the prevailing high temperatures maintain the blue component of the feature weak,  primarily \chaS{made} by \SiIII\ $\lambda$$\lambda$ 3796, 3806  and \CoIII\ $\lambda$ 3782. 
 This can be seen in Fig. \ref{fig11} where we generate a synthetic spectrum at day $-$9, deliberately excluding any calcium at high velocity at $v$ > 13,450 \kms\,. This adjustment yields negligible differences compared to the spectrum containing calcium and shows the dominance of \SiIII\ and \CoIII\,. In contrast, SN\,1999aa distinctly displays \CaII\ dominance in the same feature during the same period, an influence even more pronounced in SN\,2003du.
 
 By day $-$8, a calcium mass fraction of 0.005 is sufficient for our synthetic spectra. These values fall within the range computed for SNe 1999aa and 1991T, but are significantly lower than the values reported for the spectroscopically normal events \citep{tanaka_outermostejecta_2008, Tanaka2011}.

The presence of free electrons in the outer shells can affect the \CaII\ H\&K HV feature's strength by promoting recombination from \CaIII\ to \CaII, which \chaS{may support the presence of} hydrogen as discussed by \citet{mazzali2005HVfeatures}. This allows for various calcium and hydrogen fractions combinations to reproduce the line, as shown by \citet{Aouad_99aa}. In the case of iPTF16abc, a significant hydrogen mass fraction (up to 0.05) would be needed to noticeably change the calcium ionization state. This quantity produces observable hydrogen lines like H$\alpha$, which are absent in the spectra. Thus, the derived calcium mass fraction can be considered reliable.\\

The \CaII\ near-infrared (NIR) feature, typically observed in normal SNe\,Ia, is notably absent in the early spectra of iPTF16abc and only becomes visible a few days after maximum brightness. This behaviour resembles SNe\,1999aa and 1991T and is reproduced well in our synthetic spectra.

\begin{figure}
\includegraphics[trim={0 0 0 0},clip,width=0.5\textwidth]{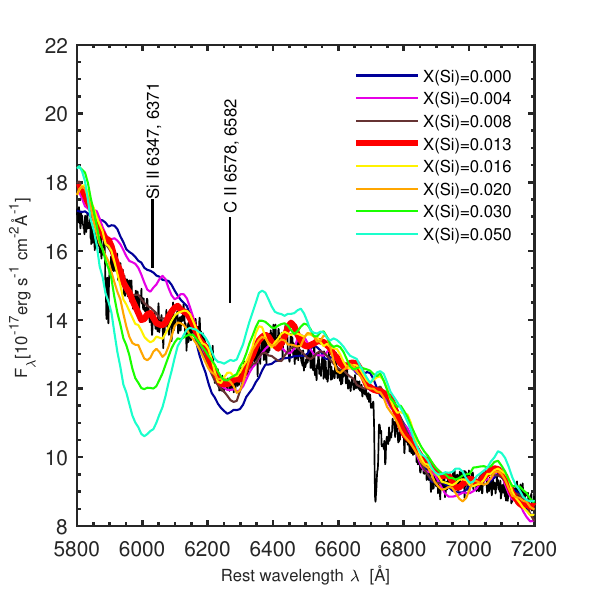}
 \caption{Probing silicon in the outer layers. The optimal replication of the \SiII\ $\lambda$ 6355 absorption line observed on day $-$~15 \chabb{with reference to $B$ maximum} is attained with a silicon mass fraction of 0.013 at velocities $v$ > 13,450 \kms. The observed spectrum is represented in black, and models with varying silicone fractions are depicted in different colors as indicated in the legend.} 
\label{fig12}
\end{figure}

\textit{Silicon, Sulphur, Magnesium:} 
The \SiII\ $\lambda$ 6355 feature is barely noticeable in the day $-$15 spectrum but gains slight strength on days $-$11 and $-$10, resembling SN\,1999aa at the same epoch. Conversely, the \SiIII\ $\lambda$ 4560 feature is strong and increases in intensity as the spectra evolve. To match these features, a Si mass fraction of 0.013 is required at $v > 13,450$ \kms, on day $-$15. Higher Si fractions would produce stronger features than observed (cf. Fig. \ref{fig12}). On day $-$11, the Si mass fraction sharply rises to 0.47 at $v > 12,200$~\kms, before gradually decreasing inward. The precise matching of the line ratios between \SiII\ and \SiIII\ features in our synthetic spectra indicates \cha{a} well-constrained ionization balance and therefore temperatures in our models.

The two \SII\ $\lambda$$\lambda$ 5468, 5654 lines begin to appear on day $-$10, and our synthetic spectra reasonably replicate their depth and evolution. The sulfur mass fraction follows a similar trend to silicon. The depletion of intermediate-mass elements in the velocity range of $\sim$ 12,000-13,000 \kms\ is akin to what has been suggested for SNe 1999aa and 1991T, contrasting with spectroscopically normal SNe\,Ia \citep{Aouad_99aa}.

The contribution of Mg to the line near 4200 \AA\ is not significant, as this line is primarily dominated by \FeIII, resembling what was observed in SNe 1999aa and 1991T. We consider a Mg mass fraction of 0.01 at $v > 13,450$ \kms.

\begin{figure}
\includegraphics[trim={0 0 0 0},clip,width=0.5\textwidth]{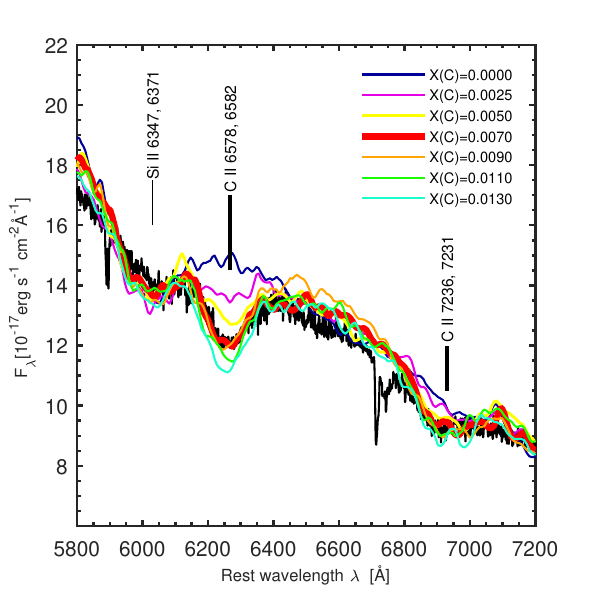}
 \caption{Probing carbon in the outermost layers. A carbon mass fraction of 0.007 provides the most accurate reproduction of the two primary \CII\ features observed on day $-$~15 \chabb{with reference to $B$ maximum} at velocities $v$ > 13,450 \kms,. (The observed spectrum is shown in black, and models with different carbon mass fractions are illustrated in various colors as indicated in the legend.)} 
\label{fig13}
\end{figure}

\begin{figure*}
 
\captionsetup[subfigure]{labelformat=empty}
	\centering
	\begin{subfigure}{1.02\textwidth}
		\includegraphics[trim={10 0 20 20},clip,width=1.02\textwidth]{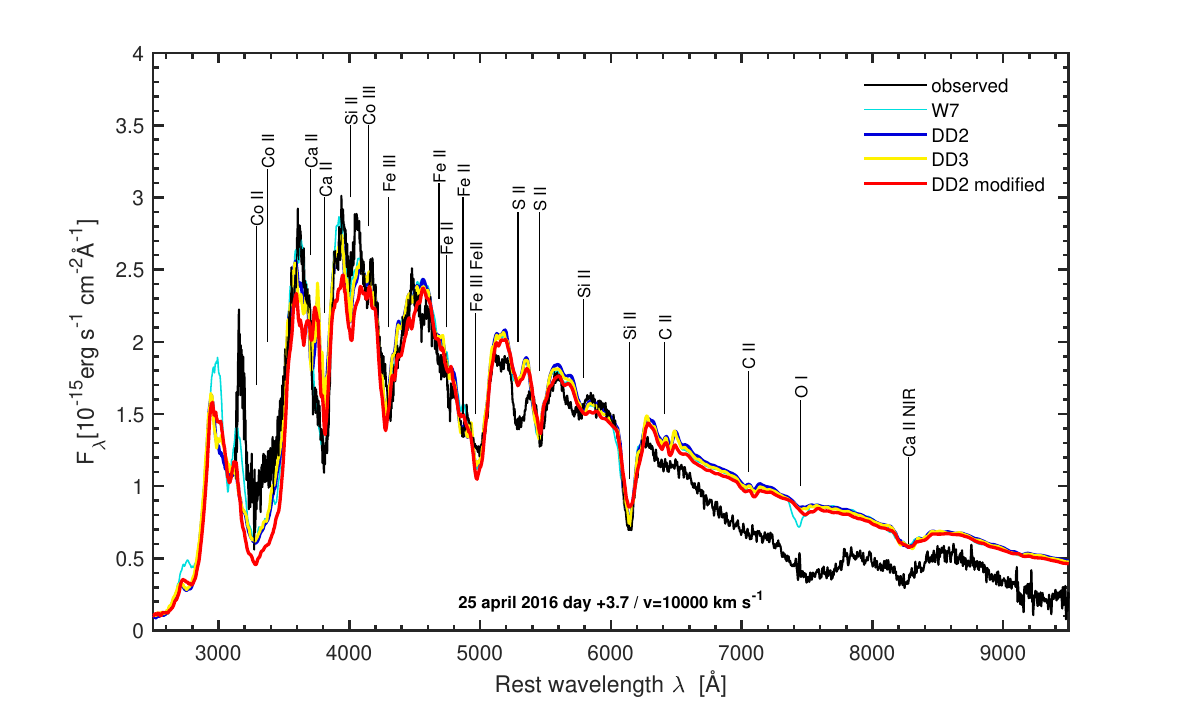}
		\caption{ }
	\end{subfigure}
	\begin{subfigure}{1.02\textwidth}
		\includegraphics[trim={10 0 20 20},clip,width=1.02\textwidth]{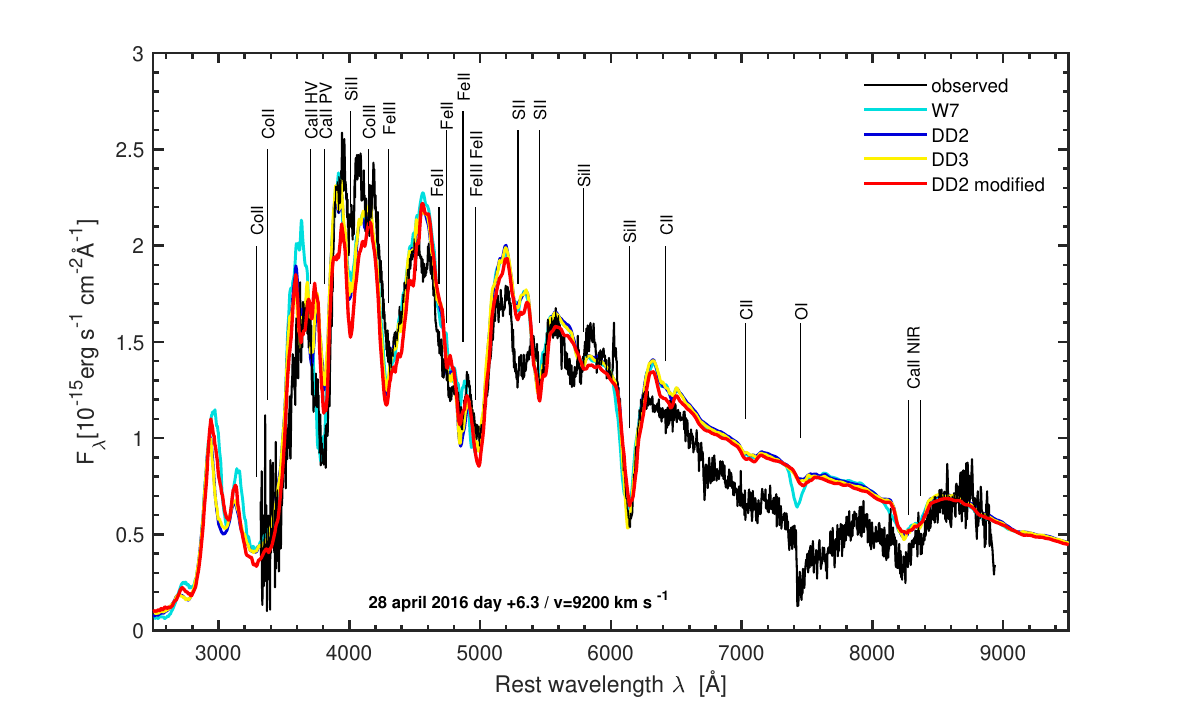}
		\caption{ }
	\end{subfigure}
 
\caption{Comparison of post-maximum observed spectra of iPTF16abc (black) with synthetic spectra in various colors, corresponding to different density profiles as specified in the legend. Features near 6780 \AA\ (weak) and
7400 \AA\ (strong) are telluric. Epochs are presented with reference to $B$ maximum. }
	
 \label{fig14}
\end{figure*}

\begin{figure*} 
\includegraphics[trim={10 0 20 10},clip,width=1.02\textwidth]{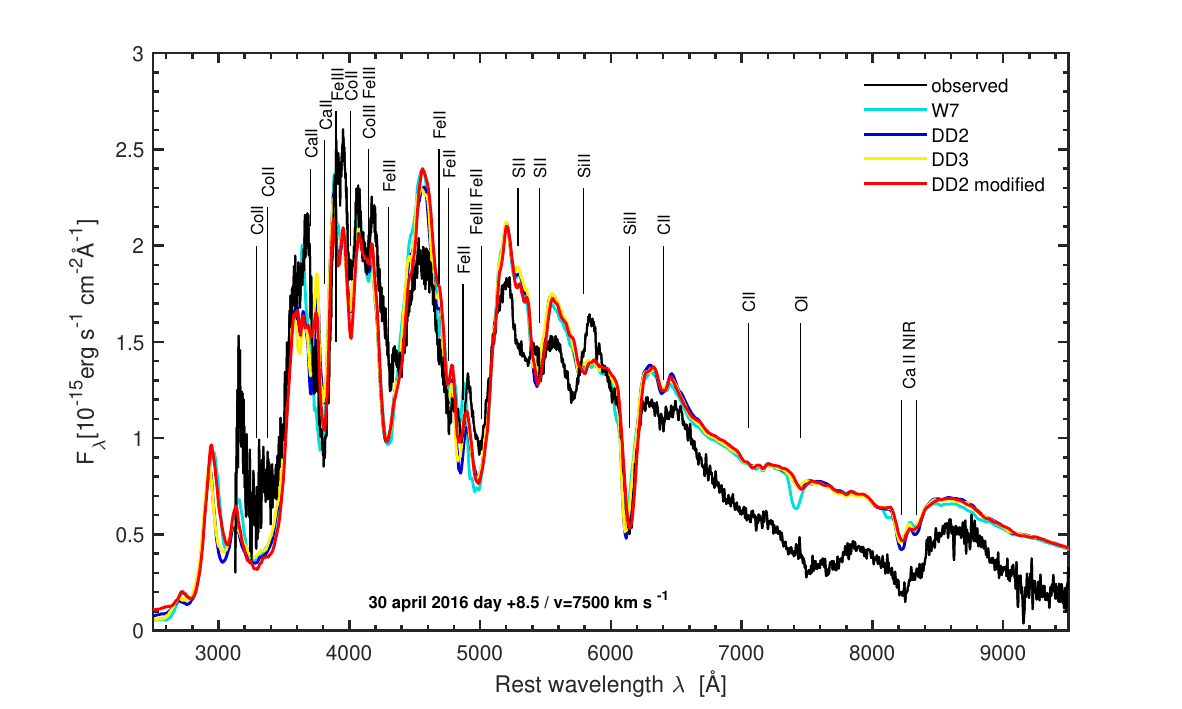}
\caption{Comparison of  day $+$ 8 observed spectrum of iPTF16abc (black) with synthetic spectra in various colors, corresponding to different density profiles as specified in the legend. Epochs are presented with reference to $B$ maximum.}  
\label{fig15}
\end{figure*}

\textit{Carbon, Oxygen:}
A strong \CII\ $\lambda$ 6580 and a weaker \CII\ $\lambda$ 7234, are evident in the day $-$15 spectrum. They diminish as \chaS{time} progresses but surprisingly re-emerge at day $+$3.7, gaining strength by day $+$8. To accurately replicate these C II features, a mass fraction of 0.007 is required at $v > 13,450$ \kms, (day $-$15, cf. Fig. \ref{fig13}) decreasing to 0.005 at $v > 11,200$ \kms\ on day $+$8.

Notably, clear carbon lines are not unambiguously present in the spectra of SNe 1999aa and 1991T. Upper limits of 0.0005 and 0.005 were  respectively reported in a velocity range of 12,500-17,000 \kms, \citep{Aouad_99aa, Sasdelli2014}, while \citet{Tanaka2011} noted a mass fraction of 0.002 for the normal SN\,2003du at $10,500 < v < 15,000$ \kms. The absence of carbon detection in SNe\,Ia may stem from observational biases such as the lack of early spectra or line blending. Nevertheless, it has been reported in a significant fraction of SNe\,Ia across all spectroscopic classifications \citep{folatelli_unburned_material,Parrentcarbonfeatures2011}.

Our synthetic spectra reproduce an \OI\ $\lambda$ 7771 line near 7400 \AA\, which is absent in the observed spectrum at day $-$15. The formation of this line has also been reported in previous works \citep{Tanaka2011, Aouad_99aa}. During this specific epoch and within this velocity range \chaS{(13,500 \kms < $v$ < 14,000 \kms)}, the density in our models drops to log$\rho \approx$-12.2 g cm$^{-3}$, while the radiation temperature reaches approximately $10^4$K. Under these conditions, the Sobolev optical depth for the \OI\ $\lambda$ 7771 line surpasses unity, leading to the formation of this spectral feature. The removal of this line from the models necessitates either substituting oxygen with alternative elements, which could introduce unobserved features in the spectra, and furthermore is difficult to explain in the context of Ia's. \chaS{Or it requires} much high temperatures, which \chaS{would} alter the ionization balance and hinder the formation of other spectral lines. Another avenue to explore is the possibility of density reduction in the outermost layers, a topic we investigate further in section \ref{sec5}.
In any case, \chab{with a mass fraction of 0.94,} oxygen prevails at $v > 13,450$\kms, but decreases inwards, as the composition becomes dominated by other elements which are needed to replicate the spectral features.

The absorption line near 6300 \AA\, has been attributed to H$\alpha$ in previous studies \citep{lentz_2002_Hydrogen_Ia}, but our synthetic spectra could not reproduce accurately this line using H, at any velocity. More importantly, replacing C with H will not \chaS{allow for the appearance of} \CII\ $\lambda$ 7234 clearly seen and accurately replicated on day $-$15.

\subsection{Post-maximum spectra}
Figures \ref{fig14} and \ref{fig15} show spectra from 3.7 to 8.5 days after $B$ maximum.
At this epoch, the photosphere resides inside the \Nifs\ dominated shell. Consequently, the assumption of a well-defined photosphere becomes less precise, and a substantial portion of the radiation energy is deposited above the photosphere \chaS{in reality}. \chaS{This, and the assumption of blackbody radiation at the photosphere may cause the excess} flux at the redder wavelengths beyond $\sim$6000 \AA\  \chaS{in the models}. However, this \chaS{supposedly} does not \chaS{strongly} affect line formation or the ionization balance in our models.

\textit{Fe-group elements:} At this epoch, several \FeII\ lines start to form and influence the main Fe dominated feature near 4800 \AA. The \FeII\ lines are accurately replicated and their ratio to \FeIII\ lines  is faithfully reproduced in our synthetic spectra, indicating  that our models are still in good ionization balance. From day $+$3.7 until day $+$6.3, the stable Fe mass fraction stabilises at 0.05, constituting $\sim$ 30 percent of the total iron content, while the remaining iron originates from the decay of \Nifs. By day $+$8.5, this fraction increases to 0.3, which is necessary to replicate the emission lines present in the nebular phase spectrum (cf. Section \ref{sec6}). \\
The \Nifs\ mass fraction remains 0.69 in the probed shells until day $+$8.5. The feature near 4000 \AA, earlier due to \SiII\ $\lambda$~4130, becomes fully dominated by \CoII\ $\lambda$ 4160 at day $+$8.5. A blend of \CoIII\ $\lambda$$\lambda$  4317, 4249 and \FeIII\ $\lambda$ 4352, create a feature near 4150 \AA\ which is also accurately replicated in our synthetic spectra. The deep feature observed near 3200 \AA\, previously dominated by \CoIII\,$\lambda$$\lambda$ 3305, 3287 is now dominated by a blend of \CoII\, lines as the temperature drops and the ionization balance shifts towards singly ionized species. 
 
\textit{Calcium:} The \CaII\ NIR appears around day $+$3.7. This feature clearly splits into two different components at day $+$6.3 and after, which is accurately replicated. The \SiIII\, and \CoIII\, contributions to the feature near 3800 \AA\ reduce with time until it is dominated by \CaII\, H\&K, with a significant contribution of \SiII\ $\lambda$ 3858 \citep{childress2014_HVfeatures}. This is expected as the temperatures drop.
 This feature is better reproduced with the W7 density profile. This may be due to the density bump in the W7 profile in the velocity shell between 13,000 \chaS{and} 16,000 \kms\ where the PV component of the line forms. The Ca mass fraction needed decreases from 0.005 at day $+$3.7 to 0.001 at day $+$8.5 at $v$ > 7500 \kms\,.

\textit{Silicon, Sulphur:} At this epoch, the prominent feature near 4400 \AA\, earlier due to \SiIII\ $\lambda$$\lambda$ 4552, 4567, 4574 becomes fully dominated by a blend of \FeIII\ $\lambda$$\lambda$  4420, 4431, 4395 and a \CoIII\ $\lambda$ 4433  lines. As a result, the two separated features at 4200 and 4400 \AA\, earlier seen in the pre-maximum epochs, blend into one deep absorption feature, which is accurately replicated. This is common to all type Ia that we have modelled \citep[cf.][]{Tanaka2011, Aouad_99aa, Sasdelli2014}. The \SiII\ $\lambda$ 6355 gets deeper as the spectra evolve. The \SiII\ $\lambda$ 5972 line appears at day $+$3.7 as depicted in our synthetic spectra. 
At day $+$6.3 the Si fraction is 0.17. At v < 7500 \kms, \Nifs\ and stable iron dominate the abundance.

\begin{figure}
\includegraphics[trim={0 10 0 22},clip,width=0.37\textwidth]{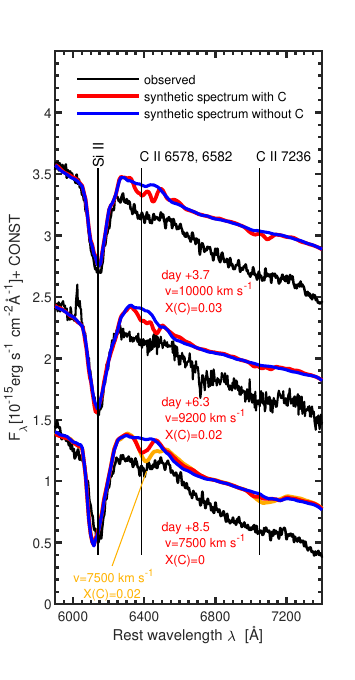}
 \caption{Probing carbon at the inner layers: spectra taken approximately 4 to 8 days after the peak. To accurately replicate the observed feature at around 6400 \AA\,, carbon presence is required down to velocities as low as $v$=9,200 \kms\,. Introducing carbon at even lower velocities would cause the absorption line to shift redder from its observed position, as depicted in the yellow synthetic spectrum generated at day $+$ 8.5. \chabb{ Epochs are presented with reference to $B$ maximum.}}  
\label{fig16}
\end{figure}

\textit{Carbon, Oxygen:} The \chab{carbon} mass fraction increases from 0.005 at $v > 11,200$\kms, to 0.03 at $v > 10,000$\kms, to slightly decrease to 0.025 at $v > 9,200$\kms. \chaS{The overall increase over the  shells between 9200 and 11800 \kms (cf. Fig. \ref{fig22})}, is essential to replicate the \CII\ $\lambda$ 6580 at days $+$3.7, $+$6.3 and $+$8.5. Notably, there is no requirement for carbon presence below 9200 \kms, as any carbon abundance at this velocity would shift the line to the red, deviating from its observed position (cf. Fig. \ref{fig16}). At day $+$8.5, the velocity of the \CII\, line is $\sim$ 8000 \kms,  500 \kms detached above the photosphere (v$_{ph}$=7,500 \kms). 
As oxygen cannot be accurately constrained from any visible line in the spectra, a similar abundance as carbon is assumed since it is difficult to justify the presence of carbon without oxygen.

\section{A modified density profile }
\label{sec5}

Our primary motivation for exploring modifications to the density profile arises from the presence of the \OI\ $\lambda$ 7771 line in our models, \chab{using DD2, DD3 and W7 density profiles}, which is absent in the data obtained from the earliest spectrum taken at day $-$15.
To address this, we experiment with modifying the original DD2 density profile. We find that the most favorable outcome is achieved by removing 0.05 M$_\odot$ of mass at v > 13,450 \kms, while simultaneously adding the missing mass in the deeper layers. This ensures the conservation of the total mass, while the total kinetic energy is not significantly affected due to the small mass shift (cf. Fig. \ref{fig4}). This approach has been already used in \citet{hachinger2009_05bl, Ashall2016} and in \citet{mazzali_2014_2011fe}. This minor modification does not substantially affect the abundance as a function of velocity neither the integrated yields. Attention has been given to avoid steep density gradients. Such gradients might trigger the formation of lines within narrower velocity ranges, thus potentially yielding lines significantly narrower than those observed.
The results are shown in Figures \ref{fig5}, \ref{fig6}, \ref{fig7}, \ref{fig14} and \ref{fig15} with red lines. Notably, the manipulation of the density profile had minimal visible impact on most lines, but it reduced the \OI\ \chaS{signature} substantially. Removing more mass from the outer layers will eventually affect the formation of other lines.

In general, all density profiles we used yield reasonably good fits to the observed spectra, which makes it difficult to favour any one model. The same result was obtained when modelling SN\,1999aa. Upon closer examination, both DD models exhibit a striking resemblance and outperform the W7 model in certain spectral regions, such as the vicinity of 3800 \AA\, during the pre-maximum epoch, whereas W7 performs better in the same region during the post maximum epoch. The W7 model consistently generates a narrow and robust \OI\ $\lambda$ 7771 line, even in post-maximum spectra. The modified DD2, despite \chaS{wiping} out the strong \OI\ $\lambda$ 7771 in the earliest spectrum, falls short in reproducing the \SiII\ $\lambda$ 6355 line in the post maximum spectra at days $+$3.7 and $+$6.3. \chaS{Also the model does not} reach the peak flux around 4000 \AA\,. Therefore, we deduce that the delayed detonation models appear to be reasonable, although they may necessitate some adjustments, in line with the conclusions drawn for SN\,1999aa \citep{Aouad_99aa}. Our inclination towards DD models is also substantiated by insights gained from the abundance distribution and the total \Nifs\ content, which we explore in greater detail in Section \ref{sec7}.

\section{THE NEBULAR PHASE}
\label{sec6}

Modelling the nebular spectrum allows us to \chab{derive} the properties of the inner ejecta. Nebular spectra of SNe\,Ia are dominated by forbidden lines of [\FeII] and [\FeIII] in emission and usually display lines of a few other elements, such as Mg and Ca \citep[\eg][]{Mazzali2008}. The emission intensity is a diagnostic for the content of \Nifs. Most \Nifs\ in SNe\,Ia is normally located at low velocity, and so the nebular phase is the best time to see it directly. At the typical epochs of nebular spectra ($\sim 1$ year), most \Nifs\ has decayed into \Fefs. 
The residual \Cofs\,still heats the gas upon decaying into \Fefs, and this is balanced by cooling via line emission. Computing the heating rate and balancing this with cooling yields an estimate of the mass of \Nifs\ synthesised in the explosion, as well as the overall mass and composition of the emitting region, which is usually confined to the region where the abundance of \Nifs\ is high. Additionally, the presence of lines of two different ions of iron, \FeII\ and \FeIII, is very useful in order to estimate the recombination rate, and hence the density of the gas. The ionization balance is affected by additional cooling provided by the stable iron atoms that are also synthesised during explosive burning if the density is sufficiently high. Estimating the mass of stable Fe is a powerful tool to estimate the mass of the progenitor white dwarf at the time of explosion, and ultimately to differentiate between different explosion scenarios, as stable Fe is thought to be synthesised only if the white dwarf has a mass close to the Chandrasekhar limit. 

Unfortunately, the nebular spectrum for iPTF16abc at day $+$342.4 has limited wavelength coverage and low signal-to-noise ratio. Furthermore, no photometric measurements are available at this epoch. In order to calibrate the spectrum, we extrapolated the $g$-band photometry measurement from a few days earlier. Additionally, we used the $U-g$ colour of a similar supernova, SN\,1999aa, as an approximation to calibrate the $U$-band. While this approach is rudimentary and susceptible to potential inaccuracies, it can still yield results that fall within $\approx 20$ percent of the true ones. Consequently, there is some uncertainty associated with the derived \Nifs\ mass. However, these quantities are subjected to further validation against the bolometric light curve, which we explore in Section \ref{sec8}.

The available nebular spectrum of iPTF16abc was modelled with our SN nebular code \citep{mazzali_2001_1998bw_nebularcode}. The code computes the deposition of the $\gamma$-rays and positrons produced by \Nifs\ and \Cofs\ decay using a \chab{Monte Carlo} scheme \citep{Cappellaro1997}, and balances the heating that this causes with cooling via emission lines in non-local thermodynamic equilibrium (NLTE), following the principles laid out in \citet{Axelrod1980}. The best-fitting density profile, WDD2, was used for our models, together with the outer layers' composition determined by photospheric-epoch spectral fitting above. The composition in the inner layers, which are only visible at late times, was determined by modelling the nebular spectrum itself. These inner layers are dominated by Fe-group elements. \Nifs\ was introduced until a flux was reached that was comparable to the observed one. Unfortunately the nebular-phase spectrum of iPTF16abc  does not extend far enough to the red for us to be able to estimate the mass of calcium, which is an important coolant despite its typically low abundance. This introduces some uncertainty into our results.

\begin{figure}
\includegraphics[trim={10 2 10 15},clip,width=0.5\textwidth]{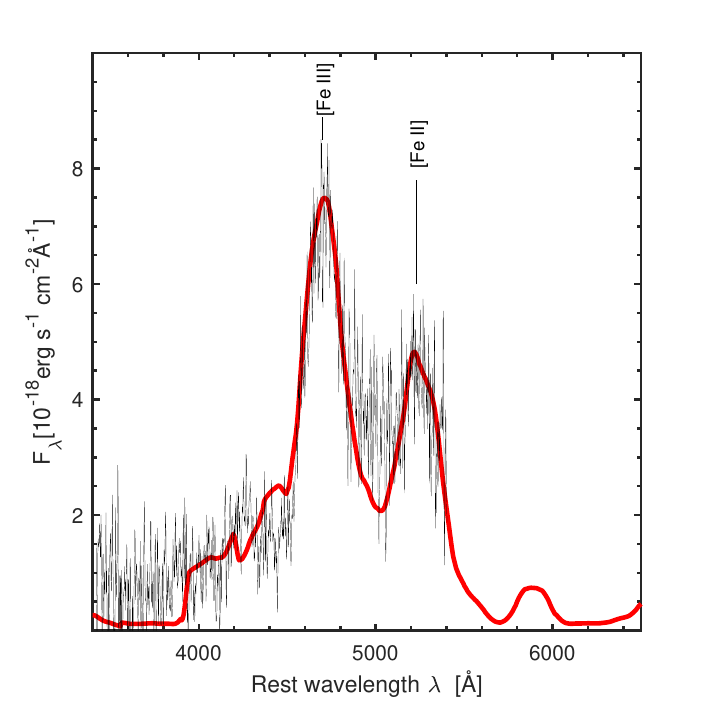}
\caption{Nebular spectrum of iPTF16abc on day $+$342 with reference to $B$ maximum light, displayed in black, compared with our synthetic spectrum in red.} 
\label{fig17}
\end{figure}

\section{ABUNDANCE TOMOGRAPHY}
\label{sec7}

\begin{figure*}
\includegraphics[trim={0 0 0 0},clip,width=0.8\textwidth]{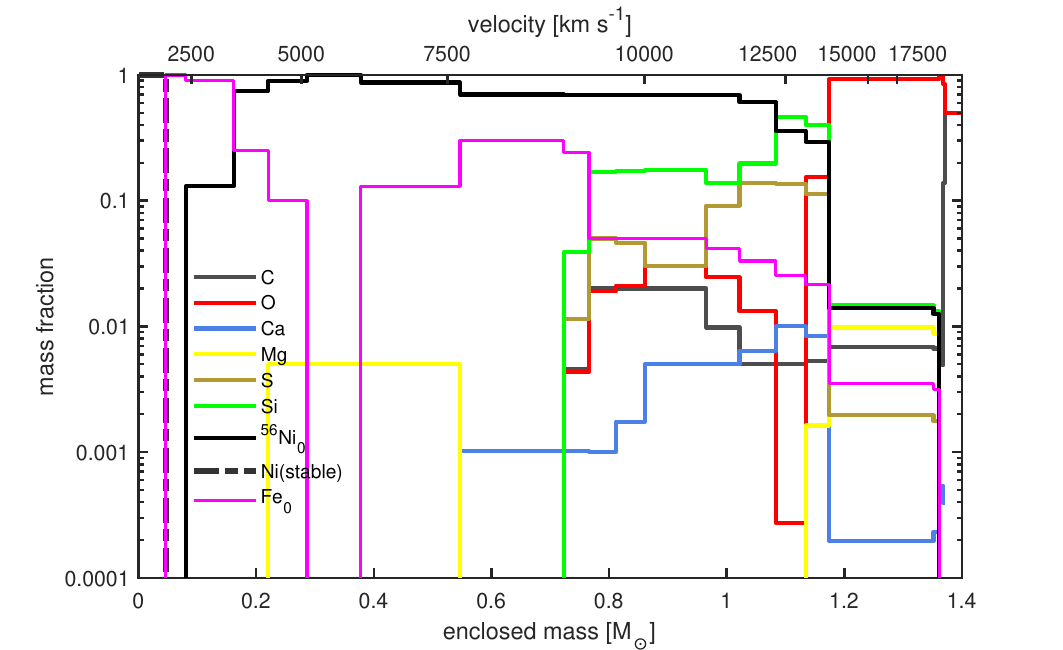}\\
\includegraphics[trim={0 0 0 0},clip,width=0.8\textwidth]{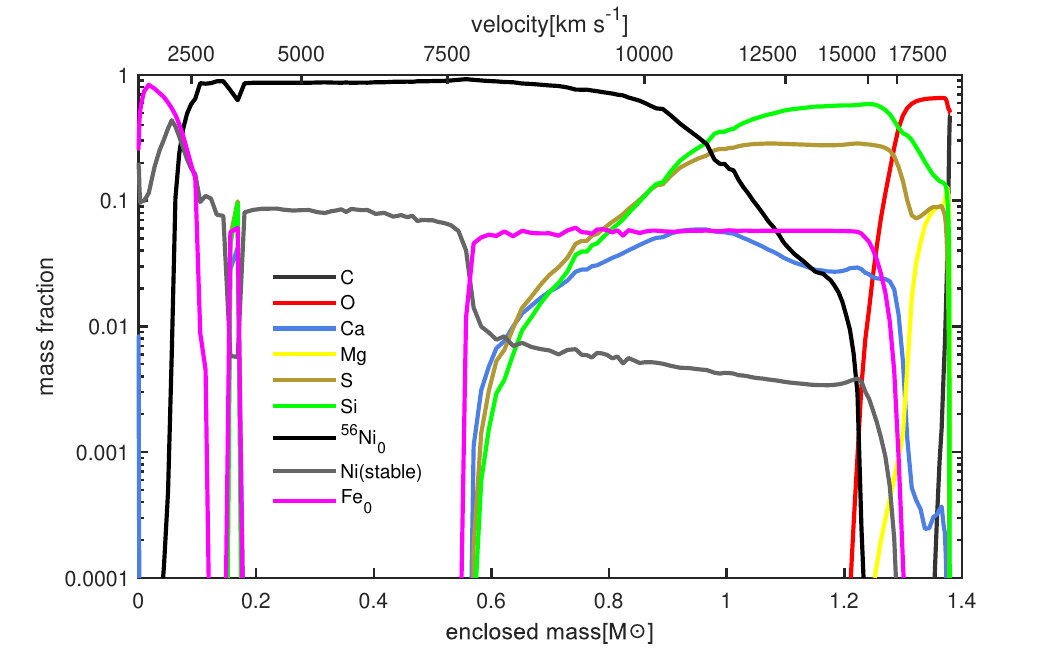}\\
\caption{Upper panel: abundances obtained from spectral models using the DD3 density profile. Lower panel: the original nucleosynthesis from DD3   \citep{iwamoto1999}.}  
\label{fig18}
\end{figure*}

Using the results of both the photospheric and the nebular phases we construct the abundance as a function of depth. This is shown in  Fig. \ref{fig18} where we plot the mass fraction of the main elements as a function of velocity and mass compared to the original abundance profile of the DD2 hydrodynamic model \citep{iwamoto1999}.

The inner core, up to about 3500 \kms, is dominated by neutron rich stable iron group elements.
The nucleus that prevails within the innermost layer is \Nife, but our calculations indicate its depletion occurring at 1000 \kms. In contrast, in the original model, \Nife\ persists much farther outward. This consistent finding was observed in all the events we modeled. \chab{Recently, stable IGEs have been confidently detected in JWST spectra of normal SNe Ia \citep{2022xkqDerkacy2023JWST}, further supporting the results of our modeling}. Situated above this \Nife-dominated layer is a layer primarily composed of stable iron (\ie\ \Feff), which extends until 3500\,\kms.

In the shell spanning from 5,000 to $\approx 12,500$\,\kms, \Nifs\ emerges as the dominant nucleus with a combined mass of $\sim$~0.76\,M$_{\odot}$, accompanied by a portion of stable iron. Unlike the high density inner shells, these outer shells remain unaffected by electron capture but instead undergo complete Si burning, aligning with the original hydrodynamic model.
As we progress outward, the density gradually decreases, hindering the achievement of complete Si burning. Consequently, the quantity of Ni gradually diminishes, making way for IMEs, with Si and S being the primary species accompanied with a portion of Ca. The abundance profile of this IMEs shell is similar to the hydrodynamical model, exhibiting a gradual increase as the abundance of \Nifs\ starts to decrease.

However, contrary to the original model, our calculations reveal that the Si abundance peaks at $\approx 12,500$\,\kms\ and experiences sudden depletion at $v > 13,400$\,\kms. Consequently, the Si-rich layer occupies a narrower velocity range than in the original model, where it extends out to $v = 17,500$\,\kms. This bears a striking resemblance to the behaviour observed in SNe\,1999aa and 1991T, suggesting a similar explosion mechanism \citep{Aouad_99aa, Sasdelli2014}. This is further supported by the abrupt increase in oxygen abundance prevailing in the outermost layers beyond 15,000 \kms.
Interestingly, carbon is present in the shells dominated by Si and unexpectedly extends and increases inwards until about $9,200$\,\kms. 
It is difficult to explain the increase of C abundance at these deep shells with conventional symmetric explosion models. 
Unburned carbon is typically expected at $v \gsim 15,000$\,\kms, in the W7 model and even higher velocities in the DD models \citep{nomoto1984, iwamoto1999}.
Diverging from established explosion models, \Nifs\, and stable iron are
both present in the outermost layers until $v = 25,000$\,\kms, albeit at percentage level. \\

\begin{figure} 
\includegraphics[trim={0 0 0 0},clip,width=0.45\textwidth]{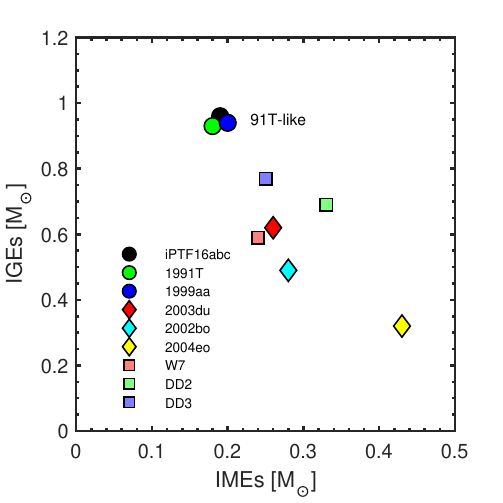}
 \caption{Scatter plot showing the total mass of Intermediate Mass Elements (IMEs) plotted against the total mass of Iron Group Elements (IGEs) for iPTF16abc as computed from our modelling, compared to other SNe and explosion models. Notably, the 91T-like events are observed to occupy a distinct region within this space }  
\label{fig19}
\end{figure}

The integrated yields \chab{from our results} are shown in Table \ref{tab2}. Using these values we compute the total kinetic energy using the formula 
\begin{multline}
    E_{k}=[1.56M (\ce{^{56}Ni}) + 1.74M(\mathrm{NSE}) + 1.24M \mathrm{(IME)}- 0.46]\\ \times 10^{51}\mathrm{ergs}
\end{multline}
\citep{woosly2007}, \chabb{where $M$(\Nifs), $M$(NSE) and $M$(IME) correspond  to the ejected mass of \Nifs, nuclear statistical equilibrium elements and IMEs respectively}. This yields a value of 1.32 $\times$ $10^{51}$ ergs.
    The \Nifs\, mass closely resembles that of the original DD3 model. However, the kinetic energy falls slightly short when compared to both the DD2 and the DD3 models (cf. Table \ref{tab2}). This outcome is expected due to the insufficient presence of IMEs, which is not adequately compensated for by an increase in \Nifs\, production. This is a common trait for all the 91T-like we have modelled. Interestingly, when we assess the mass ratio IMEs/IGEs, the 91T-like objects exhibit a common value (Table \ref{tab2} and in Fig. \ref{fig19}), which is notably lower than that of both the normal events and the hydrodynamical models. The IGEs mass of $\sim$ 0.2 \Msun\, is in the range expected for all SNe\,Ia \citep{mazzali2007}. We estimate the mass of carbon to be $\sim$ 0.01 \Msun\ above 9,200 \kms\, which aligns with the expected range of 0.001 to 0.01 \Msun, as reported by \citet{folatelli_unburned_material}.
    In general, the abundance distribution more closely resembles delayed detonation models. Explaining the extension of the \Nifs\,-dominated shell this far out in the ejecta using the W7 model proves challenging, especially considering that the total \Nifs\ mass suggested from our modelling surpasses the total \Nifs\ mass anticipated from the W7 model.
    

 \begin{table}
\setlength{\tabcolsep}{3.5pt}
\centering                                                                  \caption{Nucleosynthetic yields and kinetic energies from the modelling compared to the original hydrodynamic models. Results from other SNe are also shown. $E_\mathrm{k}$ for models is calculated using equation (1) in the text.}
\begin{threeparttable}
\scalebox{0.9}{
\hskip-1.0cm\begin{tabular}{llccccccc} 
\hline
  & \Nifs & $\mathrm{Fe}$\,$^a$ & $\mathrm{Ni}_{\mathrm{stable}}$ & IME\,$^b$  &O& IME/IGE &  $E_\mathrm{k}$ \\
  & M$_\odot$ & M$_\odot$ & M$_\odot$ & M$_\odot$ & M$_\odot$ &  & $10^{51}$\,{\rm ergs} \\
\hline
iPTF16abc (DD2)  & 0.76 & 0.20  &  0.006  & 0.19  &  0.21 &0.197 & 1.32 \\
\hline
original W7  & 0.59 & 0.16  &  0.122  & 0.24  &  0.143 &0.275 & 1.30 \\
original DD2 & 0.69 & 0.10  &  0.054  & 0.33  &  0.066 &0.39 & 1.40 \\
original DD3 & 0.77 & 0.10  &  0.0664 & 0.25  &  0.056 &0.27 & 1.43 \\
\hline
1991T (DD3) & 0.78 & 0.15  &  0.0006  & 0.18  &  0.29 &0.19 & 1.24  \\
1999aa (DD2) & 0.65 & 0.29  &  0.006  & 0.20  &  0.22 &0.21 & 1.32 \\
2003du (W7) & 0.62 & 0.18  &  0.024   & 0.26  &  0.23 &0.32 & 1.25  \\
2002bo (W7) & 0.49 & 0.27  &  0.0001  & 0.28  &  0.11 &0.37 & 1.24 \\
2004eo (W7) & 0.32 & 0.29  &  0.0005  & 0.43  &  0.3  &0.70  & 1.1 \\ 
\hline
\end{tabular}}
\begin{tablenotes}
\small      
\item $^a$All stable isotopes except for \Fefs, decay product of \Nifs 
\item $^b$ $^{28}$Si + $^{32}$S
\end{tablenotes}
\end{threeparttable}
\label{tab2}  
\end{table}

\section{BOLOMETRIC LIGHT CURVE}
\label{sec8}

We generate a bolometric light curve spanning 3000--10000\,\AA, (cf. Fig. \ref{fig20}) utilizing the $UBVri$ photometry data reported in \citet{Miller_2018_photometry_of_iptf16abc}. \chaS{This will then be compared to a light curve computed from our model (cf. Section \ref{subsec8.1})}. In cases where multiple photometric measurements were available for a given epoch, we calculate the average magnitude; this introduces an error of $\sim$ 0.1 mag. $U$ band photometry data are not available beyond day $+$22 from $B$ maximum light. Therefore, for these epochs, we assume $U$ band measurements by applying a constant $U-B$ value as computed from the last epoch where these two measurements were available (\ie day $+$21). Next, we interpolate the data at a daily resolution and convert the magnitudes to fluxes by applying the flux zero-points of \citet{FukugitaFLUXZEROPOINTS}, dereddened with the extinction curve of \citet{cardelli} using $E(B-V)$=0.08 mag \citep{Miller_2018_photometry_of_iptf16abc}. For each epoch, we compute the flux via trapezoidal integration between the specific flux values obtained at the central wavelength of each passband. We extrapolate the spectrum with a linearly decreasing function until 3000 \AA\ and 10000 \AA\ to extend the flux calculation beyond the $U$ and $i$ band central wavelengths. Finally, we compute the bolometric luminosities using the adopted distance modulus of 35.01 considering an error of 0.15 mag \citep{hyperleda}.

The supernova exhibited its highest luminosity, measuring $L_{\mathrm{peak}}=1.60 \pm 0.1 \times$ $10^{43}$ ergs s$^{-1}$, with a rise time of 18.74 days in the supernova rest frame. The bolometric peak was observed to occur roughly one day prior to $B_{\mathrm{max}}$, which is in line with previous studies examining bolometric rise time \citep{contardo2000_bolometric_risetime, scalzo_2012}.

\citet{Miller_2018_photometry_of_iptf16abc} reported a bolometric peak of 1.2 $\pm$ 0.1 $\times10^{43}$ for iPTF16abc using a distance modulus of 34.89\, $\pm$ 0.10 mag. 
We adopted a distance modulus of 35.01\,mag, consistent with our modelling. Using a shorter distance would require a smaller luminosity input, which in turn would not reproduce the ionization balance found in our models. Even if we use the distance from \citet{Miller_2018_photometry_of_iptf16abc}, it would result in a bolometric peak of $1.48 \pm 0.1 \times 10^{43}$, still more luminous than SN\,1999aa and closer to SN\,1991T, consistent with observations of their spectra. Additionally, our computed bolometric luminosities from synthetic spectra align well with both modeled and constructed light curves, serving as an additional robust validation of our method and results (see crosses in Fig. \ref{fig20}). 
Using the \Nifs\, abundance derived from our synthetic spectra (0.76 \Msun), we compute the peak luminosity using a version of Arnett rule as described in \citet{Stritzinger_Ni_Massformula_arnett}, using our rise time of 18.74 days. We find a peak luminosity of ${\mathrm{L}_{\mathrm{peak}}}=1.51 \pm\, 0.1 \times$ $10^{43}$ ergs s$^{-1}$. This result aligns with the peak luminosity value derived from our synthetic spectra yields, if we take in account the uncertainties associated with distance, photometry and the method of integration. It is worth noting that this approximation, while providing a reasonable estimate, \chaS{does not take into account} the distribution of \Nifs\ throughout the ejecta. To account for this distribution, we conduct a detailed modeling of the light curve in the subsequent subsection, offering a more accurate assessment.

\subsection{Modelling the bolometric light curve}
\label{subsec8.1}
One independent way to test the results of abundance tomography is to use the density and abundance distributions derived from spectral modelling to compute a synthetic light curve. This should in principle reproduce the observed one, if the tomography results are reasonably correct, as the light curve depends on the mass and distribution of \Nifs\ and on the opacity in the ejecta, which depends on the composition of the ejecta and governs the outwards diffusion of the photons. 

Starting from the tomography results, we used our Monte Carlo SN light curve code to compute a synthetic bolometric light curve to be compared to the one obtained from the observations. The code is based on the principles set out in \citet{Cappellaro1997} and \citet{mazzali2001}. The basic assumption is that opacity in a SN\,Ia is dominated by line opacity \citep{PauldrachNLTElineblocking}, 
and therefore it depends on composition: heavier elements such as iron have more electrons and therefore a complex energy level structure and more spectral lines that can in principle hinder photon diffusion than do lighter elements such as silicon. 
\chab{In our calculations, opacity is treated using a scheme outlined in \citet{mazzali2006Podsiadlowski},
which includes the dependence of the opacity on the number of effective lines and temperature
through the Fe mass fraction and time after the explosion, respectively.}


The deposition of the energy carried by the $\gamma$-rays and positrons is computed exactly like in the nebular code, but then the energy that has gone into heating the gas is transformed into ``optical packets'' which diffuse in the ejecta, encountering an opacity determined locally by the composition in the various ejecta layers. When a packet escapes, it is accounted as contributing to the emerging luminosity at that particular time. The code is grey in frequency, and so the output is a bolometric luminosity. This compares to the bolometric luminosity obtained from the data, but it includes radiation at all wavelengths, and so it is a ``true'' bolometric luminosity. 

\begin{figure*}
\includegraphics[trim={20 20 20 20},clip,width=0.75\textwidth]{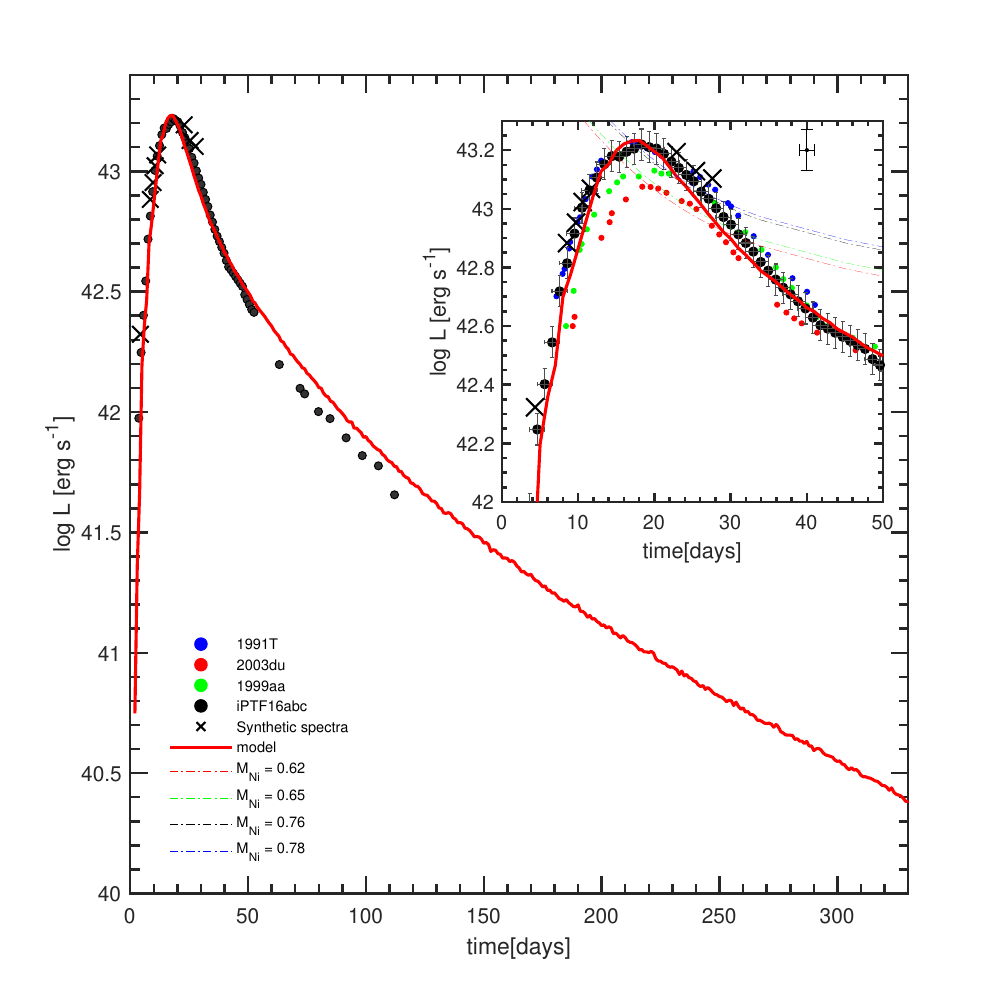}
 \caption{The modelled bolometric light curve (red line) of iPTF16abc is plotted alongside the bolometric light curves of SNe 1991T, 1999aa, and 2003du. Days are with respect to the assumed explosion time. Black dots are bolometric luminosities calculated using the photometry data of \citet{Miller_2018_photometry_of_iptf16abc}. Crosses represent bolometric luminosities calculated using our synthetic spectra. Curves in dashed lines in the subplot represent unique instantaneous energy release rates resulting from the decay of \Nifs, computed for different \Nifs\ masses for each supernova as suggested from the modelling. \citep[see][]{Sasdelli2014, Aouad_99aa, Tanaka2011}. The plotted data points include errors originating from photometry, while the error bars inserted into the plot account for distance uncertainties, considering a distance modulus error of 0.15 mag, as outlined in \citet{hyperleda}.} 
\label{fig20}
\end{figure*}

The synthetic light curve for iPTF16abc is shown in Fig. \ref{fig20}. It matches the observed one quite well in the rising part, which is  \chaS{an indication} of the correct estimate of the outer distribution of \Nifs, and near peak, which supports the mass estimate. It does a good job also in the early declining phase, while it departs from the observed light curve at later times, but clearly not at the nebular epoch, which is not shown in Fig. \ref{fig20}. These differences may be attributed to the progressive shifting of the flux towards the infrared, whose contribution is difficult to estimate because this wavelength range which is not covered by the data of iPTF16abc. It is now known that the IR becomes a strong component at late times \citep{Deckers2023_photom_NIR}.
Still, our result confirms that the estimate of the \Nifs\ mass ejected is correct, and the fact that the width of the light curve near maximum is correctly reproduced is also supportive of our estimate of \Mej\ and E$_{\mathrm{k}}$, \ie of our choice of explosion model.

\section{DISCUSSION}
\label{sec9}



\begin{figure} 
\includegraphics[trim={0 1 20 0},clip,width=0.38\textwidth]{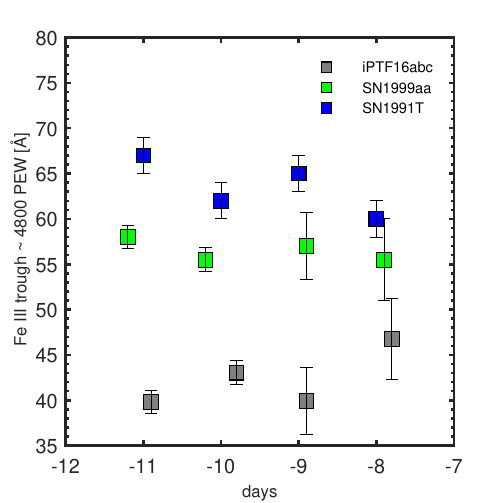}
 \caption{Comparison of the pseudo equivalent width of the \FeIII\ trough, around 4800 \AA, in the early spectra of iPTF16abc with those measured for SNe 1999aa and 1991T. The measurements were carried out following the approach described in  \citet{hachinger2006_spect_diversity_Ia}. Days are with reference to $B$ maximum.}  
\label{fig21}
\end{figure}

\begin{figure*}
\includegraphics[trim={0 0 0 0},clip,width=0.85\textwidth]{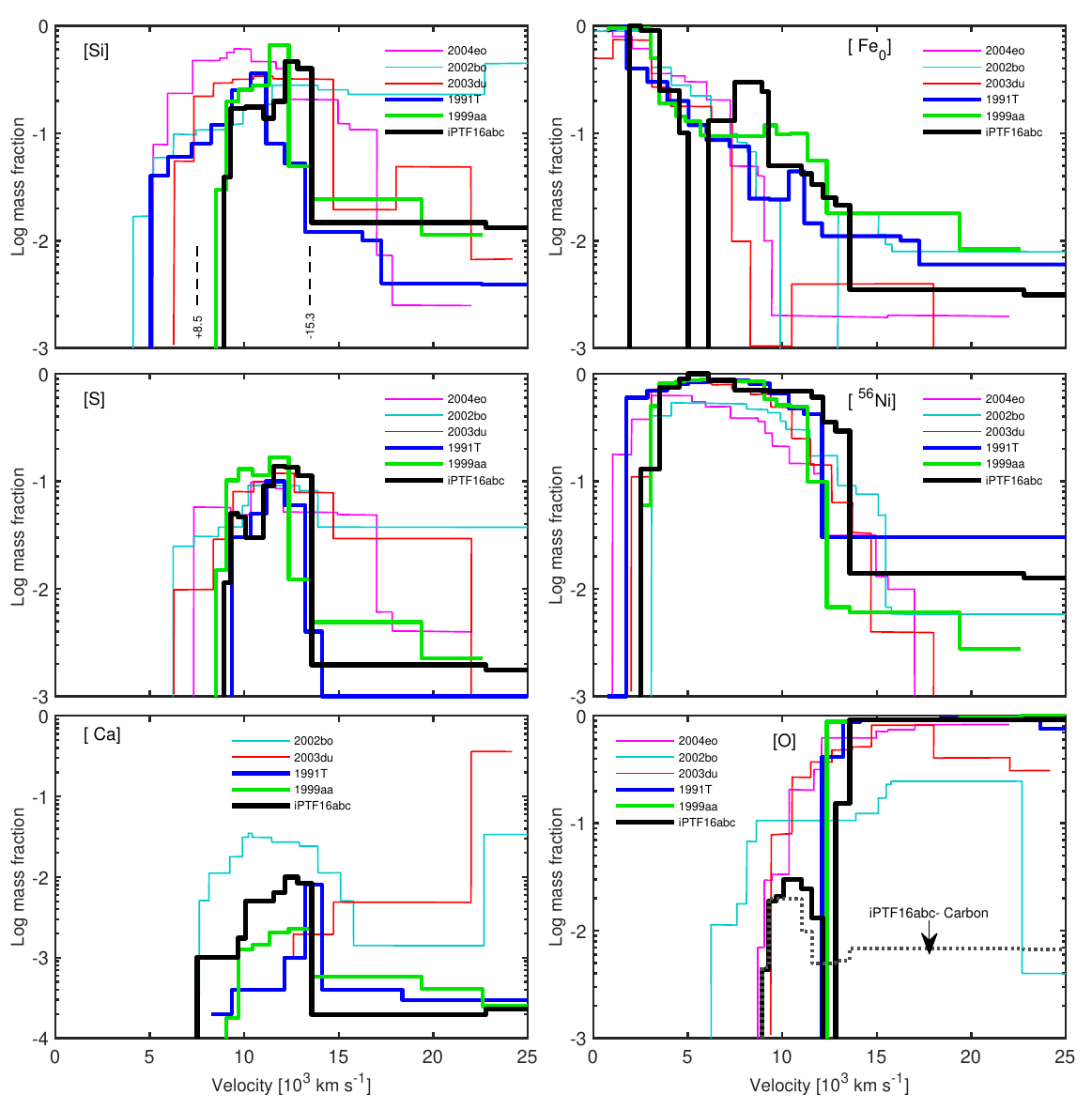}
 \caption{The distribution of the most important elements in SNe Iptf16abc, 1991T, 1999aa and some spectroscopically normal SNe Ia. Left-hand side, top to bottom: Si, S
and Ca. Right-hand side, top to bottom: stable Fe, \Nifs, O and C. Iptf16abc exhibits very similar stratification properties to SNe 1999aa and 1991T which are different than the spectroscopically normal events: the IMEs shell is narrow and is suddenly depleted above $\sim$ 12,000 \kms, oxygen dominates the outer shells and is sharply depleted below $\sim$ 12,000 \kms . Carbon extends deep and increases creating a noticeable peak around 10,000 \kms. The dashed lines in the first panel represent the velocities probed by our \cha{photospheric phase} models, spanning from approximately days $-$15.3 to $+$8.5 relative to $B$ maximum, while the regions below are probed by the nebular phase model.}  
\label{fig22}
\end{figure*}

iPTF16abc demonstrates weaker \FeIII\ lines in its early spectra compared to both SNe 1999aa and 1991T. This difference is evident in Fig. \ref{fig21} where we measure the pseudo equivalent width (PEW) of the \FeIII\, trough around 4800 \AA\, following the method described in \citet{hachinger2006_spect_diversity_Ia}. This trend is further confirmed when we examine the abundance levels (cf. Fig. \ref{fig22}). Notably, iPTF16abc exhibits approximately 7-10 times less iron than both SNe 1999aa and 1991T at velocities larger than 15,000 \kms. However, this disparity diminishes at deeper layers, where iPTF16abc's iron abundance aligns more closely with that of other events. This is supported by PEW measurements of iron emission lines in the nebular phase (see \citet{Miller_2018_photometry_of_iptf16abc}). Consequently, this suggests that the relative weakness of \FeIII\ lines during the photospheric phase is likely due to variations in abundance within the outermost shells rather than being driven by ionization-related luminosity effects.\\ 
This observation is particularly significant, highlighting that objects with weaker early \FeIII\ lines may have comparable or even higher luminosities than those with stronger lines like SNe 1999aa and 1991T.
This comparison becomes particularly intriguing when examining the 2003fg-like events. Despite their elevated luminosities, this sub-class is recognized by weak \FeIII\, lines \citep{ashall_2021_2003fg_superchandra}. 
 Moreover, the outermost shells' supersolar abundance of stable iron in iPTF16abc, coupled with the existence of \Nifs, is a topic of interest. While the presence of \Nifs, is undeniably linked to explosive nucleosynthesis, we should also remain receptive to the possibility that the presence of stable iron might be connected to the progenitor star's original metallicity. This alternative explanation should not be dismissed, especially in cases where the abundance is supersolar, as observed in SN\,1999aa \citep{Aouad_99aa}, SN\,1991T \citep{mazzali1995, Sasdelli2014}  SN\,2002bo \citep{Stehle2005}, SN\,2010jn \citep{Hachinger2013-2010jn}, and iPTF16abc. Such instances may suggest that progenitors with above-solar metallicity do not necessarily lead to dimmer events, as suggested, for instance, in \citet{timmes2003_metallicity}.  
\chab{An alternative explanation may also be multidimensional effects involving bubbles of burned material rising to outer regions, or large scale turbulent mixing giving rise to the presence of IGEs at the outer shells \citep{Hachisu_1992_mixing, Aspden_2010_turbulent_mixing}.}

\begin{figure*} 
	\begin{tabular}{ll}
    \includegraphics[trim={0 1 10 0},clip,width=0.4\textwidth]{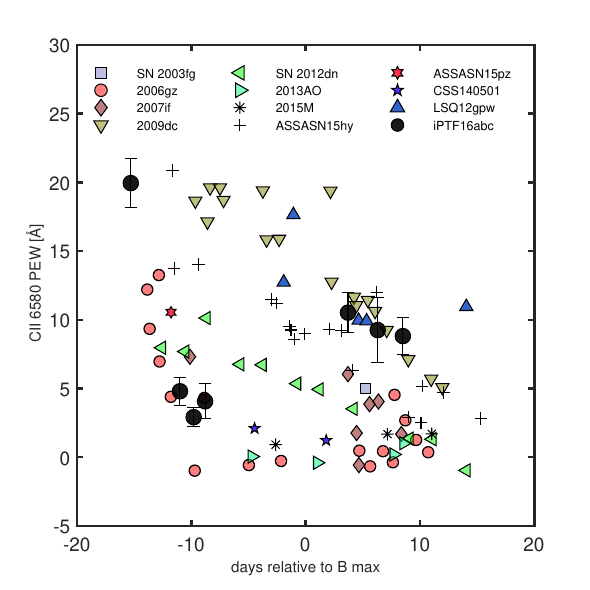}&\includegraphics[trim={0 0 10 0},clip,width=0.4\textwidth]{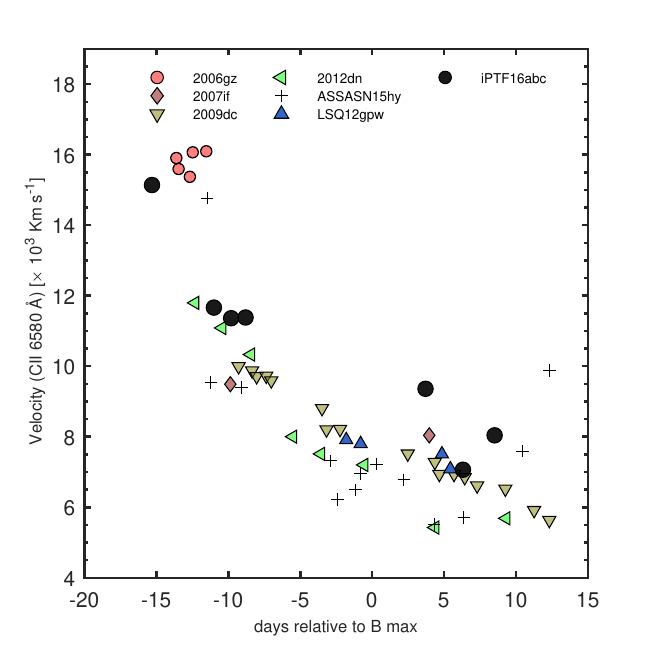}\\

    \end{tabular}
    \caption{Left panel: Pseudo equivalent width (PEW) of the \CII\, 6580 \AA\, observed in  iPTF16abc, compared to a set of SNe 2003fg-like taken from \citet{ashall_2021_2003fg_superchandra}. The line initially shows notable strength at an early stage but experiences a subsequent weakening before ultimately strengthening again, approximately one week after maximum light.\\
    Right panel: Velocity of the same line}  
    \label{fig25}
	\end{figure*}

 When examining the spectral evolution from the spectroscopically normal to the peculiar luminous 91T-like events, the attenuation of the \SiII\ $\lambda$ 6355 line can be more attributed to abundance-related factors rather than influences of ionization \citep{Sasdelli2014, Aouad_99aa}. 
Interestingly, upon comparing the abundance profiles of different events (cf. Fig. \ref{fig22}), silicon displays a noticeable decline in the outermost shells across the three 91T-like objects we have modeled, within a comparable velocity range. A similar observation has been recently proposed by \citet{obrien2023_1991TlikeSNe}, where they model the optical spectra of 40 SNe\,Ia encompassing both normal and 91T-like classifications. They deduce a deficiency in intermediate-mass elements (IMEs) abundance in the 91T-like events compared to those classified as normal \citep[see also][]{mazzali2007}. This behavior holds true for sulfur abundance as well. This trend implies the existence of a distinct burning regime within this specific subclass of supernovae, setting it apart from the dynamics that characterize spectroscopically normal events like SN\,2003du.
Within the operative regime among the 91T-like events, it appears that the burning front abruptly ceases within a velocity range spanning between 12,000 and 13,000 \kms. This cessation creates a division within the outer ejecta, resulting in two distinct regions characterized by differing degrees of burning activity, one marked by vigorous burning and the other characterized by subdued burning.\\

The persistent carbon features until two weeks post maximum display no resemblance to any SN\,Ia within the same spectroscopic subclass, namely SNe 1999aa and 1991T. Moreover, despite sporadic appearances of these attributes in the early spectra of events classified as normal, they are notably absent during the phase of maximum brightness and subsequent periods \eg SN2013\,du \citep{zheng20113du}, SN\,2011fe \citep{nugent2011fe}
\citep[see also][]{folatelli_unburned_material, Parrentcarbonfeatures2011}.
Remarkably, these characteristics have consistently emerged in 2003fg-like events \citep{ashall_2021_2003fg_superchandra}, such as SN 2009dc \citep{taubenberger_2011_2009dc, hachinger_2009dc}, SN 2006gz \citep{hicken_2006gz}, SN 2012dn \citep{chakdrahari_2012dn} and recently SN\,2022pul \citep{siebert2023_2022pul_photopheric}. This pattern is evident from early spectra up to two weeks after reaching peak intensity.\\
In Fig. \ref{fig25} we plot the PEW of the \CII\ $\lambda$ 6580 and its velocity compared to a set of 2003fg-like SNe taken from \citet{ashall_2021_2003fg_superchandra}. While the velocity of the line follows a similar trend to that observed in 2003fg-like events, the PEW displays a different pattern. It initially exhibits considerable strength at an early stage, experiences subsequent weakening, but unexpectedly strengthens again one week after maximum. This behavior distinguishes it from other 03fg-like supernovae and reinforces the insights derived from our abundance analysis, that reveals an increase in carbon abundance in deeper layers, giving rise to a "bump" around 10,000 \kms. Given this observation, it is plausible to speculate that this phenomenon arises from mixing driven by flame propagation asymmetries or clumps of unburned material along the line of sight \citep{folatelli_unburned_material, Parrentcarbonfeatures2011}. 
Alternatively, it may arise from a C-rich envelope, leading to the decrease of photospheric velocities as the ejecta interacts with this extended envelope. This interaction could give rise to the prominence of carbon lines and longer diffusion times, a scenario suggested for the 2003fg-like events \citep{ashall_2021_2003fg_superchandra}. This scenario may be considered viable for iPTF16abc given the low velocity of the \SiII\, 6355 \AA\ line  \citep[see]{Miller_2018_photometry_of_iptf16abc}.
Finally, carbon could be the surviving material from a pulsation driven detonation (PDD) \citep{ivanova_PDD_1974, Khokhlov1991_DD_PDD, dessart2014pdd}. \chaS{This is} a scenario in which a failed deflagration expands the WD, creating a low density loosely bound envelope, such that, when the detonation occurs, an outer unburnt shell is left and the IME shell is confined in a narrow velocity range.

Regardless of the scenario, the identification of carbon in iPTF16abc is captivating, positioning this event as a link connecting the 03fg-like and 91T-like classifications. Nonetheless, in addition to these spectroscopic parallels with 03fg-like events, iPTF16abc distinctively retains all the spectroscopic characteristics reminiscent of a 91T-like SN\,Ia. 
This resemblance to the 91T-like SNe\,Ia persists into the nebular phase. Furthermore, when assessing the photometric attributes, iPTF16abc bears no resemblance to 03fg-like events, which are characterized by IR light curves devoid of a secondary maximum \citep{ashall_2021_2003fg_superchandra}. Additionally, the time of the i-band maximum occurs after to the B-band maximum in the events classified as 03fg-like. In iPTF16abc it occurs three days before the B band maximum, similar to the majority of 91T-like events \citep[see][]{Ashall2020_SBV_tib}.

The subdued intensity of the \CaII\ H\&K feature in the early-phase spectra of iPTF16abc, compared to SN\,1999aa, can easily be attributed to the distinct luminosities of the two objects. However, drawing definitive conclusions becomes challenging when juxtaposed with SN\,1991T, particularly given the comparable bolometric light curves exhibited by these events, even when considering inherent uncertainties. The observed dissimilarities in the behavior of the \CaII\ H\&K lines do not offer sufficient grounds to categorize iPTF16abc as an intermediate object between SN\,1999aa and SN\,1991T. Based on the available data, it is more plausible to consider iPTF16abc as a similar object to SN\,1991T, except for the Carbon features.
The differences in the suppression of the \CaII\ H\&K line could potentially be influenced by density variations in the outer layers affecting calcium recombination, or it could arise from different ionization levels due to varying temperatures.



It remains an open question whether SNe displaying intermediate properties \chaS{between the ones observed in iPTF16abc and the ones observed in both SNe 1999aa and 1991T} will be discovered, particularly concerning the strengths and temporal evolution of the \FeIII, \CaII, H\&K, \CII\ and \SiII\ features. This could provide further insights into our understanding of the complex spectroscopic diversity observed in \chaS{91T-like} SNe\,Ia.

\section{CONCLUSIONS }
\label{sec10}

We have modelled a time series of optical spectra to derive the abundance of the peculiar carbon rich, 91T-like type\,Ia supernova, iPTF16abc and compare it to both SNe, 1999aa and 1991T from one side, and the normal SN\,2003du from the other side.
Our main conclusions can be summarized as follows:

\begin{itemize}
\setlength{\itemsep}{5pt}

\item Delayed detonation density profiles align \chaS{very well} with iPTF16abc's spectroscopic features, even though requiring minor adjustments in the outer shells. 

\item The inner composition is dominated by stable IGEs (0.20 \Msun) as in all SNe\,Ia followed by a \Nifs\, dominated shell with a total mass of 0.76 \Msun\, and a thin IME shell (0.18 \Msun). 


\item Iron abundance variations in the outermost layers explain the early weak \FeIII\ lines, highlighting that 91T-like objects with weaker \FeIII\ lines are not necessarily dimmer or cooler than those with stronger lines.

\item Detection of stable iron in the outermost shells, similar to SNe 1999aa and 1991T challenges existing explosion models. The idea that these are caused by bubble of burned material rising to the outer layers is contradicted by the simultaneous presence of carbon \cha{within the \Nifs\ dominated shell.} This may suggest a potential link to the progenitor's original metallicity and may indicate that progenitors with above solar metallicity, may not necessarily lead to dimmer events.

\item Similar to SNe 1999aa and 1991T, the early \SiII\ $\lambda$ 6355 weakening results from silicon depletion in the outer shells. This distinct burning regime ends around 12,000 \kms, leaving incompletely burned outer layers \chaS{and} setting it apart from typical events.

\item Carbon co-exists with IGEs in the outermost shells, extending inward with a noticeable increase in deeper layers. This persistence of the carbon feature beyond the maximum light epoch establishes a unique connection with SN 2003fg-like events.


\end{itemize}

\section{ACKNOWLEDGMENTS}
\chab{C.A acknowledges support from NASA grants JWST-GO-02114, JWST-GO-02122 and JWST-GO-04522. Support for programs \#2114, \#2122, and \#4522 were provided by NASA through a grant from the Space Telescope Science Institute, which is operated by the Association of Universities for Research in Astronomy, Inc., under NASA contract NAS 5-03127.} 
The authors express their gratitude to Suhail Dhawan for sharing the spectra in readable format, \chabb{and to Adam Miller for supplying the photometry data. Additionally, the authors extend their thanks to the anonymous referee for their meticulous review and valuable suggestions, which enhanced the overall quality of the paper.}

\section*{Data availability}
The spectroscopic data underlying this article are available at the Weizmann Interactive Supernova Data Repository (WISeREP) \citep{wiserep}.






\bibliographystyle{mnras}
\bibliography{references}









\bsp	
\label{lastpage}
\end{document}